# The MIMO Iterative Waterfilling Algorithm

Gesualdo Scutari, *Member, IEEE,* Daniel P. Palomar, *Member, IEEE,* and Sergio Barbarossa, *Member, IEEE*

*Abstract*—This paper considers the non-cooperative maximization of mutual information in the vector Gaussian interference channel in a fully distributed fashion via game theory. This problem has been widely studied in a number of works during the past decade for frequency-selective channels, and recently for the more general MIMO case, for which the state-of-the art results are valid only for nonsingular square channel matrices. Surprisingly, these results do not hold true when the channel matrices are rectangular and/or rank deficient matrices.

The goal of this paper is to provide a complete characterization of the MIMO game for *arbitrary* channel matrices, in terms of conditions guaranteeing both the uniqueness of the Nash equilibrium and the convergence of asynchronous distributed iterative waterfilling algorithms. Our analysis hinges on new technical intermediate results, such as a new expression for the MIMO waterfilling projection valid (also) for singular matrices, a mean-value theorem for complex matrix-valued functions, and a general contraction theorem for the multiuser MIMO watefilling mapping valid for *arbitrary* channel matrices. The quite surprising result is that uniqueness/convergence conditions in the case of tall (possibly singular) channel matrices are more restrictive than those required in the case of (full rank) fat channel matrices.

We also propose a modified game and algorithm with milder conditions for the uniqueness of the equilibrium and convergence, and virtually the same performance (in terms of Nash equilibria) of the original game.

## I. INTRODUCTION

THE interference channel is a mathematical model relevant to many communication systems where multiple uncoordinated links share a common communication medium (see, e.g., [1]-[3]). In this paper, we focus on the *MIMO Gaussian interference channel* and consider the non-cooperative maximization of mutual information on each link in a fully distributed fashion using a game theoretical approach. The system is modeled as a (strategic non-cooperative) game where every MIMO link is a player who competes against the others by choosing his transmit covariance matrix (transmission strategy) to maximize his own information rate (payoff function), given constraints on the transmit power and treating the interference generated by the other users as additive noise (implying that multiuser encoding/decoding and interference cancellation techniques are not considered). Based on the celebrated notion of Nash Equilibrium (NE) in game theory (see, e.g., [4], [5]), an equilibrium for the whole system is reached when every player's reaction is "unilaterally optimal", i.e., when, given the rival players' current strategies, any change in a player's own strategy would result in a rate loss.

Since the seminal paper by Yu et al. [6] in 2002 (and the conference version in 2001), this problem has been studied in a number of works during the past seven years for the case of *SISO frequency-selective channels* or, equivalently, a set of parallel non-interfering scalar channels [7]-[15]. Several sufficient conditions have been derived that guarantee the uniqueness of the NE and the convergence of different distributed waterfilling based algorithms: synchronous sequential [6]-[9], [12], [13], [15], synchronous simultaneous [10]-[13], [15] and asynchronous [12]. See [16] for a unified view of the state-of-the-art results in SISO frequency-selective interference channels.

The more general MIMO case, as considered in this paper, is a nontrivial extension of the SISO frequency-selective case. There are indeed only a few papers that have studied (special cases of) the MIMO game [16]-[21]. In [17], the authors focused on the two-user MISO flat-fading channel. In [18]-[20], the authors considered the rate maximization game in MIMO interference channels, but they provided only numerical results to support the existence of a NE of the game. Furthermore, in these papers there is no study of the uniqueness of the equilibrium and convergence of the proposed algorithms. In [21], the authors showed that the MIMO rate maximization game is a concave game (in the sense of [22]), implying the existence of a NE for any set of arbitrary channel matrices [22, Theorem 1]. As far as the uniqueness of the equilibrium is concerned, only [21] showed that if the channel matrices are full column-rank and the multiuser interference is almost negligible, then the NE is unique, without quantifying how small the interference must be. Finally, in [16], the authors provided sufficient conditions that, differently from [21], explicitly quantify how strong the multiuser interference can be to guarantee both the uniqueness of the equilibrium and the convergence of distributed algorithms. However, the results in [16] are valid only for the MIMO game with *square nonsingular* channel matrices. Hence, the study of the game in the case of *arbitrary* channel matrices has not been done to date.

The goal of this paper is to fill this gap and provide a complete analysis of the MIMO game, without making any restrictive assumption on the structure of the channel matrices, so that they can be rectangular and possibly rank deficient. Interestingly, the study of the game in this more general setup cannot be obtained by a trivial extension of the framework developed in [16], since the key results used in [16] to derive conditions for both the uniqueness of the NE and the convergence of distributed algorithms–the interpretation

G. Scutari and D. P. Palomar are with the Department of Electronic and Computer Engineering, Hong Kong University of Science and Technology, Hong Kong. E-mails: <eealdo,palomar>@ust.hk. This work was supported by the NSFC/RGC 2009/11 research grant.
S. Barbarossa ais with the INFOCOM Department, University of Rome, "La Sapienza,"Rome, Italy. E-mail: sergio@infocom.uniroma1.it.
Manuscript received July 16, 2008; revised October 14, 2008.

of the MIMO waterfilling solution as a projection of a proper *nonsingular* matrix and the contraction theorem of the MIMO multiuser waterfilling mapping−fail to hold in the more general case of rectangular and possibly rank deficient channel matrices. Thus, the study of the game hinges on new technical intermediate results, such as a new expression for the MIMO waterfilling projection valid (also) for singular matrices and a mean-value theorem for complex matrix-valued functions, which will play a key role in deriving one on the main results of the paper: a general contraction theorem for the multiuser MIMO watefilling mapping that is valid for *arbitrary* channel matrices. A quite surprising result follows from this new theorem: the sufficient conditions guaranteeing both the uniqueness of the fixed point of the MIMO waterfilling mapping−the Nash equilibrium of the MIMO game−and the convergence of totally asynchronous iterative waterfilling algorithms (IWFA) in the case of (strict) tall and/or singular channel matrices need to be stronger than those required for the case of full row-rank (and thus also square) channel matrices.

To provide a weaker set of uniqueness and convergence conditions for both cases of tall and fat channel matrices, we propose a new best-response strategy for each player based on a modified game. The resulting new game admits a unique NE under a unified set of sufficient conditions that are valid for arbitrary channel matrices and coincide with the weakest ones obtained for the original game (i.e., those obtained in the case of full row-rank channel matrices). Such conditions are also sufficient for the convergence of an asynchronous distributed algorithm based on this new mapping. Numerical experiments show that the performance of both games is virtually the same.

The paper is organized as follows. Section II gives the system model and formulates the optimization problem as a strategic non-cooperative game. In Section III, we derive the main properties of the multiuser MIMO waterfilling mapping: a new expression of the MIMO waterfilling solution as a matrix projection and the contraction theorem for the multiuser MIMO waterfilling mapping, both valid for arbitrary channel matrices. The contraction property of the multiuser waterfilling paves the way to derive sufficient conditions guaranteeing the uniqueness of the fixed point of the waterfilling projector−alias the NE of the MIMO game−and the convergence of iterative, possibly asynchronous, distributed algorithms as detailed in Sections IV and V, respectively. Section VI provides some numerical results validating our theoretical findings, and Section VII draws some conclusions.

The following notation is used in the paper. Uppercase and lowercase boldface denote matrices and vectors respectively. The operators $(\cdot)^*$, $(\cdot)^H$, $(\cdot)^\sharp$, $\mathcal{E}\{\cdot\}$, $\text{vec}(\cdot)$ and $\text{Tr}(\cdot)$ are conjugate, Hermitian, Moore-Penrose pseudoinverse [27], expectation, stacking vectorization operator and trace operator, respectively. The range space and null space operators are denoted by $\mathcal{R}(\cdot)$ and $\mathcal{N}(\cdot)$, respectively. The operators $\leq$ and $\geq$ for vectors and matrices are defined component-wise, while $\mathbf{A} \succeq \mathbf{B}$ (or $\mathbf{A} \preceq \mathbf{B}$) means that $\mathbf{A} - \mathbf{B}$ is positive (or negative) semidefinite. The operator $\text{Diag}(\cdot)$ is the diagonal matrix with the same diagonal elements as the matrix (or vector) argument, $[\cdot]_{ij}$ denotes the $(i,j)$ element of the matrix argument, $\otimes$ denotes the Kronecker product operator [26], $\oplus$ denotes the direct sum [26], and $(x)^+ \triangleq \max(0, x)$. The spectral radius[1] of a matrix $\mathbf{A}$ is denoted by $\rho(\mathbf{A})$. The operator $\mathbf{P}_{\mathcal{N}(\mathbf{A})}$ (or $\mathbf{P}_{\mathcal{R}(\mathbf{A})}$) denotes the orthogonal projection onto the null space (or the range space) of matrix $\mathbf{A}$ and it is given by $\mathbf{P}_{\mathcal{N}(\mathbf{A})} = \mathbf{N}_A(\mathbf{N}_A^H\mathbf{N}_A)^{-1}\mathbf{N}_A^H$ (or $\mathbf{P}_{\mathcal{R}(\mathbf{A})} = \mathbf{R}_A(\mathbf{R}_A^H\mathbf{R}_A)^{-1}\mathbf{R}_A^H$), where $\mathbf{N}_A$ (or $\mathbf{R}_A$) is any (strictly tall) matrix whose columns are linear independent vectors spanning $\mathcal{N}(\mathbf{A})$ (or $\mathcal{R}(\mathbf{A})$) [29], [30]. The operator $[\mathbf{X}]_{\mathscr{Q}} = \operatorname{argmin}_{\mathbf{Z} \in \mathscr{Q}} \|\mathbf{Z} - \mathbf{X}\|_F$ denotes the matrix projection with respect to the Frobenius norm[2] of matrix $\mathbf{X}$ onto the (convex) set $\mathscr{Q}$. The sets $\mathbb{C}$, $\mathbb{R}$, $\mathbb{R}_+$, $\mathbb{R}_{++}$, $\mathbb{N}_+$, and $\mathbb{S}_+^{n \times n}$ (or $\mathbb{S}_{++}^{n \times n}$) stand for the set of complex, real, nonnegative real, positive real, nonnegative integer numbers, and $n \times n$ complex positive semidefinite (or definite) matrices, respectively.

## II. THE RATE MAXIMIZATION GAME

We consider a vector Gaussian interference channel, composed of $Q$ MIMO links. The transmission over the generic $q$-th MIMO channel with $n_{T_q}$ transmit and $n_{R_q}$ receive dimensions can be described by the baseband signal model

$$\mathbf{y}_q = \mathbf{H}_{qq}\mathbf{x}_q + \sum_{r \neq q} \mathbf{H}_{rq}\mathbf{x}_r + \mathbf{n}_q, \quad (1)$$

where $\mathbf{x}_q \in \mathbb{C}^{n_{T_q}}$ is the vector transmitted by source $q$, $\mathbf{H}_{qq} \in \mathbb{C}^{n_{R_q} \times n_{T_q}}$ is the direct channel of link $q$, $\mathbf{H}_{rq} \in \mathbb{C}^{n_{R_q} \times n_{T_r}}$ is the cross-channel matrix between source $r$ and destination $q$, $\mathbf{y}_q \in \mathbb{C}^{n_{R_q}}$ is the vector received by destination $q$, and $\mathbf{n}_q \in \mathbb{C}^{n_{R_q}}$ is a zero-mean circularly symmetric complex Gaussian noise vector with arbitrary covariance matrix $\mathbf{R}_{n_q}$ (assumed to be nonsingular). The second term on the right-hand side of (1) represents the Multi-User Interference (MUI) received by the $q$-th destination and caused by the other active links. For each transmitter $q$, the total average transmit power is

$$\mathcal{E}\left\{\|\mathbf{x}_q\|_2^2\right\} = \text{Tr}(\mathbf{Q}_q) \leq P_q, \quad (2)$$

where $\mathbf{Q}_q \triangleq \mathcal{E}\{\mathbf{x}_q\mathbf{x}_q^H\}$ is the covariance matrix of the transmitted vector $\mathbf{x}_q$, and $P_q$ is the maximum average transmitted power in units of energy per transmission.

Since our goal is to find distributed algorithms that require neither a centralized control nor coordination among the links, we focus on transmission techniques where no interference cancellation is performed and the MUI is treated as additive colored noise from each receiver. Each channel is assumed to change sufficiently slowly to be considered fixed during the whole transmission, so that the information theoretical results are meaningful. Moreover, perfect channel state information at both transmitter and receiver sides of each link is assumed;[3] each receiver is also assumed to measure with no errors the covariance matrix of the noise plus MUI generated by the other users.

---

[1] The spectral radius $\rho(\mathbf{A})$ of the matrix $\mathbf{A}$ is defined as $\rho(\mathbf{A}) \triangleq \max\{|\lambda| : \lambda \in \sigma(\mathbf{A})\}$, with $\sigma(\mathbf{A})$ denoting the spectrum of $\mathbf{A}$ [26].

[2] The Frobenius norm $\|\mathbf{X}\|_F$ of $\mathbf{X}$ is defined as $\|\mathbf{X}\|_F \triangleq (\text{Tr}(\mathbf{X}^H\mathbf{X}))^{1/2}$ [26].

[3] Note that each user $q$ is only required to known his own channel $\mathbf{H}_{qq}$, but not the channels $\{\mathbf{H}_{rq}\}_{r \neq q}$ of the others.

Under these assumptions, invoking the capacity expression for the single user Gaussian MIMO channel—achievable using random Gaussian codes by all the users—the maximum information rate on link $q$ for a given set of users' covariance matrices $\mathbf{Q}_1, \ldots, \mathbf{Q}_Q$ is [24]

$$R_q(\mathbf{Q}_q, \mathbf{Q}_{-q}) = \log \det \left( \mathbf{I} + \mathbf{H}_{qq}^H \mathbf{R}_{-q}^{-1}(\mathbf{Q}_{-q}) \mathbf{H}_{qq} \mathbf{Q}_q \right) \quad (3)$$

where

$$\mathbf{R}_{-q}(\mathbf{Q}_{-q}) \triangleq \mathbf{R}_{n_q} + \sum_{r \neq q} \mathbf{H}_{rq} \mathbf{Q}_r \mathbf{H}_{rq}^H \quad (4)$$

is the MUI-plus-noise covariance matrix observed by user $q$, $\mathbf{Q}_{-q} \triangleq (\mathbf{Q}_r)_{r \neq q}$ is the set of all users' covariance matrices, except the $q$-th one.

Given the above setup, we formulate the system design within the framework of game theory using as desirability criterion the concept of Nash equilibrium (NE) (see, e.g., [4], [5]). Specifically, we consider the following strategic noncooperative game:

$$(\mathscr{G}) : \quad \begin{array}{c} \text{maximize} \\ \mathbf{Q}_q \\ \text{subject to} \quad \mathbf{Q}_q \in \mathscr{Q}_q, \end{array} \quad R_q(\mathbf{Q}_q, \mathbf{Q}_{-q}) \quad \forall q \in \Omega, \quad (5)$$

where $\Omega \triangleq \{1, \ldots, Q\}$ is the set of players (i.e., the links), $R_q(\mathbf{Q}_q, \mathbf{Q}_{-q})$ is the payoff function of player $q$ defined in (3), and $\mathscr{Q}_q$ is the set of admissible strategies (the covariance matrices) of player $q$, defined as[4]

$$\mathscr{Q}_q \triangleq \left\{ \mathbf{Q} \in \mathbb{C}^{n_{T_q} \times n_{T_q}} : \mathbf{Q} \succeq \mathbf{0}, \; \text{Tr}\{\mathbf{Q}\} = P_q \right\}. \quad (6)$$

In game $\mathscr{G}$, each player $q$ competes against the others by choosing the transmit covariance matrix $\mathbf{Q}_q$ (i.e., his strategy) that maximizes his own information rate $R_q(\mathbf{Q}_q, \mathbf{Q}_{-q})$, as given in (3), subject to the average transmit power constraint in (6). A solution of the game—a NE—is reached when each user, given the strategy profiles of the others, does not get any rate increase by unilaterally changing his own strategy (see, e.g., [4], [5], [16]).

To write the Nash equilibria of game $\mathscr{G}$ in a convenient form, we first introduce the following notations and definitions. Given $\mathscr{G}$ and $\mathbf{Q}_{-q} \in \mathscr{Q}_{-q} \triangleq \mathscr{Q}_1 \times \ldots \times \mathscr{Q}_{q-1} \times \mathscr{Q}_{q+1} \times \ldots \times \mathscr{Q}_Q$, we write the eigendecomposition of $\mathbf{H}_{qq}^H \mathbf{R}_{-q}^{-1}(\mathbf{Q}_{-q}) \mathbf{H}_{qq}$ for each $q \in \Omega$ as:

$$\mathbf{H}_{qq}^H \mathbf{R}_{-q}^{-1}(\mathbf{Q}_{-q}) \mathbf{H}_{qq} = \mathbf{U}_q \mathbf{D}_q \mathbf{U}_q^H, \quad (7)$$

where $\mathbf{U}_q = \mathbf{U}_q(\mathbf{Q}_{-q}) \in \mathbb{C}^{n_{T_q} \times r_q}$ is a (semi-)unitary matrix with the eigenvectors, $\mathbf{D}_q = \mathbf{D}_q(\mathbf{Q}_{-q}) \in \mathbb{R}_{++}^{r_q \times r_q}$ is a diagonal matrix with $r_q \triangleq \text{rank}(\mathbf{H}_{qq}^H \mathbf{R}_{-q}^{-1}(\mathbf{Q}_{-q}) \mathbf{H}_{qq}) = \text{rank}(\mathbf{H}_{qq})$ positive eigenvalues, and $\mathbf{R}_{-q}(\mathbf{Q}_{-q})$ is defined in (4).

Given $q \in \Omega$ and $\mathbf{Q}_{-q} \in \mathscr{Q}_{-q}$, the solution to the single-user optimization problem in (5) is the well-known waterfilling solution (e.g., [24]):

$$\mathbf{Q}_q^\star = \mathbf{WF}_q(\mathbf{Q}_{-q}), \quad (8)$$

[4]Observe that, in the definition of $\mathscr{Q}_q$ in (6) we omitted the redundant condition $\mathbf{Q} = \mathbf{Q}^H$, since any *complex* positive semidefinite matrix must be necessarily Hermitian [26, Sec. 7.1]. Furthermore, there is no loss of generality in considering in (6) the power constraint with equality rather than inequality, since at the optimum to each problem in (5), the constraint must be satisfied with equality.

with the waterfilling operator $\mathbf{WF}_q(\cdot)$ defined as

$$\mathbf{WF}_q(\mathbf{Q}_{-q}) \triangleq \mathbf{U}_q \left( \mu_q \mathbf{I} - \mathbf{D}_q^{-1} \right)^+ \mathbf{U}_q^H, \quad (9)$$

where $\mathbf{U}_q = \mathbf{U}_q(\mathbf{Q}_{-q})$ and $\mathbf{D}_q = \mathbf{D}_q(\mathbf{Q}_{-q})$ are the eigenvectors and diagonal matrices of the eigenvalues in (7), and $\mu_q$ is the water-level chosen to satisfy $\text{Tr}\left\{ (\mu_q \mathbf{I} - \mathbf{D}_q^{-1})^+ \right\} = P_q$.

Using (8), we can now characterize the Nash Equilibria of the game $\mathscr{G}$ in a compact way as the following waterfilling fixed-point equation:

$$\mathbf{Q}_q^\star = \mathbf{WF}_q(\mathbf{Q}_{-q}^\star), \quad \forall q \in \Omega. \quad (10)$$

*Remark 1 - Related works:* The matrix nature of game $\mathscr{G}$ and the *arbitrary* structure of the channel matrices make the analysis of the game quite complicated and none of the results in the literature [6]-[13] and [21], [16] can be successfully applied to $\mathscr{G}$. The main difficulty in the analysis of the solution in (10) comes from the fact that the optimal eigenvector matrix $\mathbf{U}_q^\star = \mathbf{U}_q(\mathbf{Q}_{-q}^\star)$ of each user $q$ depends, in general, on the strategies $\mathbf{Q}_{-q}^\star$ of all the other users, through a very complicated implicit relationship—the eigendecomposition of the equivalent channel matrix $\mathbf{H}_{qq}^H \mathbf{R}_{-q}^{-1}(\mathbf{Q}_q^\star) \mathbf{H}_{qq}$.

The goal of this paper is to provide a complete analysis of game $\mathscr{G}$, without making any assumption on the structure of the channel matrices. Surprisingly, in the case of rectangular channel matrices, most of the key results given in [16],[5] such as the interpretation of the MIMO multiuser waterfilling as a matrix projection onto a convex set [16, Lemma 6] and the contraction theorem for the multiuser MIMO waterfilling operator [16, Th. 5], fail to hold, implying that a new framework is needed for the more general case of rectangular, possibly singular, channel matrices. We focus on this issue in the next section.

III. PROPERTIES OF THE MIMO WATERFILLING MAPPING

In this section we derive some interesting properties of the multiuser MIMO waterfilling mapping. These results will be instrumental to study the uniqueness of the NE of game $\mathscr{G}$ and to derive conditions for the convergence of the algorithms proposed in Section V. Differently from previous works in the literature [6]-[16], where only channel matrices with special structure—nonsingular diagonal [6]-[13], nonsingular Toeplitz circulant [14], [15], and nonsingular MIMO square [16] matrices—were considered, here we do not make any assumption on the structure of the channel matrices, which can be rectangular and (possibly) rank deficient matrices. The main result of the section is a contraction theorem for the multiuser MIMO waterfilling mapping, valid for arbitrary channel matrices.

*A. MIMO waterfilling as a projector*

In [15], we showed that the waterfilling operator for SISO frequency-selective channels can be interpreted as a projection onto a simplex set. This gave us a key tool to prove the convergence of the iterative waterfilling based algorithms in the multiuser case. This interpretation can be naturally generalized to the MIMO case for square and nonsingular channel matrices [16, Lemma 6]: given the nonsingular matrix $\mathbf{H}_{qq}^H \mathbf{R}_{-q}^{-1}(\mathbf{Q}_{-q})\mathbf{H}_{qq}$, the MIMO waterfilling operator $\mathbf{WF}_q(\mathbf{Q}_{-q})$ in (9) can be equivalently written as

$$\mathbf{WF}_q(\mathbf{Q}_{-q}) = \left[ -\left(\mathbf{H}_{qq}^H \mathbf{R}_{-q}^{-1}(\mathbf{Q}_{-q})\mathbf{H}_{qq}\right)^{-1} \right]_{\mathscr{Q}_q}, \quad (11)$$

where $\mathscr{Q}_q$ is defined in (6).

[5]Note that, because of the space limitation, most of the results in [16] dealing with the MIMO case were given without proof.

*Negative result #1:* One could conjecture that the more general case where the channel matrix $\mathbf{H}_{qq}^H \mathbf{R}_{-q}^{-1}(\mathbf{Q}_{-q})\mathbf{H}_{qq}$ is singular follows naturally from (11) by replacing the inverse with some generalized inverse [27]. Unfortunately, this conclusion is not true. In the general case of possibly singular channel matrices, the equivalent expression of the MIMO waterfilling operator as a projection contains an additional term—the orthogonal projection matrix onto the null space of the direct channel matrix—as stated next.

*Lemma 1:* The MIMO waterfilling operator $\mathbf{WF}_q(\mathbf{Q}_{-q})$ in (9) can be equivalently written as

$$\mathbf{WF}_q(\mathbf{Q}_{-q}) = \left[ -\left(\left(\mathbf{H}_{qq}^H \mathbf{R}_{-q}^{-1}(\mathbf{Q}_{-q})\mathbf{H}_{qq}\right)^{\sharp} + c_q \mathbf{P}_{\mathcal{N}(\mathbf{H}_{qq})}\right) \right]_{\mathscr{Q}_q}, \quad (12)$$

where $c_q$ is a positive constant satisfying $c_q \geq c_q(\mathbf{Q}_{-q}) \triangleq P_q + \max_{i \in \{1,\ldots,r_q\}}[\mathbf{D}_q(\mathbf{Q}_{-q})]_{ii}^{-1}$, and $\mathscr{Q}_q$ is defined in (6). An upper bound of $c_q(\mathbf{Q}_{-q})$ independent on $\mathbf{Q}_{-q}$ is given in (93) in Appendix A.

*Proof:* See Appendix A. ∎

Observe that, for each $q \in \Omega$, $\mathbf{P}_{\mathcal{N}(\mathbf{H}_{qq})}$ in (12) depends only on the channel matrix $\mathbf{H}_{qq}$ (through the right singular vectors of $\mathbf{H}_{qq}$ corresponding to the zero singular values) and not on the strategies of the other users, since $\mathbf{R}_{-q}(\mathbf{Q}_{-q})$ is positive definite for all $\mathbf{Q}_{-q} \in \mathscr{Q}_{-q}$.

*Remark 2 - Special cases:* Since the expression in (12) is valid for any arbitrary set of channel matrices, it contains as special cases previous results obtained in [15] and [16] when the channel matrices have a particular structure. For example, since $\mathbf{P}_{\mathcal{N}(\mathbf{H}_{qq})} = \mathbf{0}$ if and only if $\mathbf{H}_{qq}$ is full column-rank, in the case of square nonsingular channel $\mathbf{H}_{qq}$, the MIMO waterfilling projection in (12) coincides with (11) [16].

*Non-expansive property of the waterfilling operator:* Thanks to the interpretation of the MIMO waterfilling in (9) as a projector, one can obtain the following non-expansive property of the waterfilling operator [16, Lemma 7].

*Lemma 2:* The matrix projection $[\cdot]_{\mathscr{Q}_q}$ onto the convex set $\mathscr{Q}_q$ defined in (6) satisfies the following non-expansive property:

$$\left\| [\mathbf{X}]_{\mathscr{Q}_q} - [\mathbf{Y}]_{\mathscr{Q}_q} \right\|_F \leq \|\mathbf{X} - \mathbf{Y}\|_F, \quad \forall\, \mathbf{X}, \mathbf{Y} \in \mathbb{C}^{n_{T_q} \times n_{T_q}}. \quad (13)$$

□

*Nash equilibria as fixed-points of the WF projection:* Using (12) in (10), it is straightforward to see that all the Nash equilibria of game $\mathscr{G}$ can be alternatively obtained as the fixed-points of the mapping defined in (12):

$$\mathbf{Q}_q^\star = \left[ -\left(\left(\mathbf{H}_{qq}^H \mathbf{R}_{-q}^{-1}(\mathbf{Q}_{-q}^\star)\mathbf{H}_{qq}\right)^{\sharp} + c\mathbf{P}_{\mathcal{N}(\mathbf{H}_{qq})}\right) \right]_{\mathscr{Q}_q}, \forall q \in \Omega, \quad (14)$$

for sufficiently large (positive) $c$. This equivalent expression of the Nash equilibria along with the non-expansive property of the multiuser waterfilling, will play a key role in the study of uniqueness of the equilibrium and convergence of the distributed iterative algorithms proposed in Section V.

### B. Contraction properties of the multiuser MIMO waterfilling mapping

Building on the interpretation of the waterfilling operator as a projector, we provide now one of the main results of the paper: the contraction property of the MIMO multiuser waterfilling mapping for *arbitrary channel matrices*. To this end, we introduce first some basic definitions and results that will be used in our derivations.

*1) Intermediate definitions and miscellaneous results:* Given the multiuser waterfilling mapping

$$\mathbf{WF}(\mathbf{Q}) = (\mathbf{WF}_q(\mathbf{Q}_{-q}))_{q \in \Omega} : \mathscr{Q} \mapsto \mathscr{Q}, \quad (15)$$

where $\mathscr{Q} = \mathscr{Q}_1 \times \cdots \times \mathscr{Q}_Q$, $\mathscr{Q}_q$ and $\mathbf{WF}_q(\mathbf{Q}_{-q})$ are defined in (6) and (12), respectively, we introduce the following block-maximum norm on $\mathbb{C}^{n \times n}$, with $n = n_{T_1} + \ldots + n_{T_Q}$, defined as [23]

$$\|\mathbf{WF}(\mathbf{Q})\|_{F,\text{block}}^{\mathbf{w}} \triangleq \max_{q \in \Omega} \frac{\|\mathbf{WF}_q(\mathbf{Q}_{-q})\|_F}{w_q}, \quad (16)$$

where $\|\cdot\|_F$ is the Frobenius norm and $\mathbf{w} \triangleq [w_1, \ldots, w_Q]^T > \mathbf{0}$ is any positive weight vector. Let $\|\cdot\|_{\infty,\text{vec}}^{\mathbf{w}}$ be the *vector* weighted maximum norm, defined as [26]

$$\|\mathbf{x}\|_{\infty,\text{vec}}^{\mathbf{w}} \triangleq \max_{q \in \Omega} \frac{|x_q|}{w_q}, \quad \text{for} \quad \mathbf{w} > \mathbf{0}, \quad \mathbf{x} \in \mathbb{R}^Q, \quad (17)$$

and let $\|\cdot\|_{\infty,\text{mat}}^{\mathbf{w}}$ denote the *matrix* norm induced by $\|\cdot\|_{\infty,\text{vec}}^{\mathbf{w}}$, given by [26]

$$\|\mathbf{A}\|_{\infty,\text{mat}}^{\mathbf{w}} \triangleq \max_q \frac{1}{w_q} \sum_{r=1}^{Q} |[\mathbf{A}]_{qr}| w_r, \quad \text{for} \quad \mathbf{A} \in \mathbb{R}^{Q \times Q}. \quad (18)$$

Given the set $\mathscr{P} \triangleq (\mathbf{P}_{rq})_{r \neq q}$, with each $\mathbf{P}_{rq} \in \mathbb{C}^{n_{R_q} \times n_{R_q}}$, we introduce the nonnegative matrix $\mathbf{S}(\mathscr{P}) \in \mathbb{R}_+^{Q \times Q}$ defined as

$$[\mathbf{S}(\mathscr{P})]_{qr} \triangleq \begin{cases} \rho\left(\mathbf{H}_{rq}^H \mathbf{P}_{rq}^H \mathbf{H}_{qq}^{\sharp H} \mathbf{H}_{qq}^{\sharp} \mathbf{P}_{rq} \mathbf{H}_{rq}\right), & \text{if } r \neq q, \\ 0, & \text{otherwise.} \end{cases} \quad (19)$$

For the sake of notation, we use the following convention: when each $\mathbf{P}_{rq} = \mathbf{I}$ (i.e., $\mathscr{P} = (\mathbf{I})$), then we denote $\mathbf{S}((\mathbf{I}))$ by $\mathbf{S}$, i.e.,

$$[\mathbf{S}]_{qr} \triangleq [\mathbf{S}((\mathbf{I}))]_{qr} = \begin{cases} \rho\left(\mathbf{H}_{rq}^H \mathbf{H}_{qq}^{\sharp H} \mathbf{H}_{qq}^{\sharp} \mathbf{H}_{rq}\right), & \text{if } r \neq q, \\ 0, & \text{otherwise.} \end{cases} \quad (20)$$

Finally, given the following definition of interference-plus-noise to noise ratio:

$$\text{innr}_q \triangleq \frac{\rho\left(\mathbf{R}_{n_q} + \sum_{r \neq q} P_r \mathbf{H}_{rq} \mathbf{H}_{rq}^H\right)}{\lambda_{\min}(\mathbf{R}_{n_q})} \geq 1, \quad q \in \Omega, \quad (21)$$

let define $\mathbf{S}^{\text{up}} \in \mathbb{R}_+^{Q \times Q}$ as

$$[\mathbf{S}^{\text{up}}]_{qr} \triangleq \begin{cases} \text{innr}_q \cdot \rho\left(\mathbf{H}_{rq}^H \mathbf{H}_{rq}\right) \rho\left(\mathbf{H}_{qq}^{\sharp H} \mathbf{H}_{qq}^{\sharp}\right), & \text{if } r \neq q, \\ 0, & \text{otherwise.} \end{cases} \quad (22)$$

We provide now a couple of lemmas that will be used in the forthcoming derivations (we omit the proof because of the space limitation).

*Lemma 3:* (*Reverse order law for Moore-Penrose pseudoinverses*0) Let $\mathbf{H} \in \mathbb{C}^{m \times n}$ with $\text{rank}(\mathbf{H}) = m$ (*i.e., full rank fat matrix*), and $\mathbf{R} \in \mathbb{C}^{n \times n}$ with $\text{rank}(\mathbf{R}) = n$. Then,

$$\left(\mathbf{H}^H \mathbf{R} \mathbf{H}\right)^{\sharp} = \mathbf{H}^{\sharp} \mathbf{R}^{-1} \mathbf{H}^{\sharp H}. \quad (23)$$

□

Interestingly, a generalization of Lemma 3 that is valid also for generalized inverses (cf. [27]) of the product of matrices can be found, e.g., in [28, Th. 2.2], under some conditions.

*Negative result #2*: In the case of (strictly) *full column-rank* matrix $\mathbf{H}$, the reverse order law in (23) is not satisfied anymore (the necessary and sufficient conditions given in [28, Th. 2.2] are not satisfied). In fact, in such a case, invoking the matrix version of the Kantorovich inequality [31, Ch. 11], one can prove that the following relationship exists between the LHS and RHS in (23):

$$\left(\mathbf{H}^H \mathbf{R} \mathbf{H}\right)^{\sharp} \preceq \mathbf{H}^{\sharp} \mathbf{R}^{-1} \mathbf{H}^{\sharp H}. \quad (24)$$

This undesired result is one of the main reasons why the derivations obtained in [16], under the assumption of *square nonsingular* channel matrices (for which Lemma 3 holds true), cannot be generalized to rectangular and/or rank deficient matrices. We thus need new results, as given in Section III-B3.

Finally, we need the following.

*Lemma 4:* Let $\mathbf{X} = \mathbf{X}^H \in \mathbb{C}^{n \times n}$ and $\mathbf{A} \in \mathbb{C}^{m \times n}$. Then,

$$\left\|\mathbf{A}\mathbf{X}\mathbf{A}^H\right\|_F \leq \rho\left(\mathbf{A}^H \mathbf{A}\right) \|\mathbf{X}\|_F. \quad (25)$$

□

We can now focus on the contraction properties of the multiuser MIMO waterfilling operator. We will consider only the case where all the direct channel matrices are either full row-rank or full column-rank, without loss of generality (w.l.o.g.). The rank deficient case in fact can be cast into the full column-rank case by a proper transformation of the original rank deficient channel matrices into a lower-dimensional full column-rank matrices, as shown in Appendix C (see also Section IV).

*2) Case of full row-rank (fat/square) channel matrices:* We start assuming that the channel matrices $\{\mathbf{H}_{qq}\}_{q \in \Omega}$ are full row-rank. The contraction property of the waterfilling mapping is given in the following theorem, which generalizes [16, Th. 5].

*Theorem 5 (Contraction property of $\mathbf{WF}$ mapping):*
Suppose that $\text{rank}(\mathbf{H}_{qq}) = n_{R_q}$, $\forall q \in \Omega$ (*i.e., full rank fat/square matrices*). Then, for any given $\mathbf{w} \triangleq [w_1, \ldots, w_Q]^T > \mathbf{0}$, the mapping $\mathbf{WF}$ defined in (15) is Lipschitz continuous on $\mathscr{Q}$:

$$\left\|\mathbf{WF}(\mathbf{Q}^{(1)}) - \mathbf{WF}(\mathbf{Q}^{(2)})\right\|_{F,\text{block}}^{\mathbf{w}} \leq \|\mathbf{S}\|_{\infty,\text{mat}}^{\mathbf{w}} \\ \times \left\|\mathbf{Q}^{(1)} - \mathbf{Q}^{(2)}\right\|_{F,\text{block}}^{\mathbf{w}}, \quad (26)$$

$\forall \mathbf{Q}^{(1)}, \mathbf{Q}^{(2)} \in \mathscr{Q}$, where $\|\cdot\|_{F,\text{block}}^{\mathbf{w}}$, $\|\cdot\|_{\infty,\text{mat}}^{\mathbf{w}}$ and $\mathbf{S}$ are defined in (16), (18) and (20), respectively. Furthermore, if the following condition is satisfied

$$\|\mathbf{S}\|_{\infty,\text{mat}}^{\mathbf{w}} < 1, \quad \text{for some } \mathbf{w} > \mathbf{0}, \quad (27)$$

then, the mapping $\mathbf{WF}$ is a *block-contraction with modulus* $\beta = \|\mathbf{S}\|_{\infty,\text{mat}}^{\mathbf{w}}$.

*Proof:* Given $\mathbf{Q}^{(1)} = \left(\mathbf{Q}_q^{(1)}, \ldots, \mathbf{Q}_Q^{(1)}\right) \in \mathscr{Q}$ and $\mathbf{Q}^{(2)} = \left(\mathbf{Q}_1^{(2)}, \ldots, \mathbf{Q}_Q^{(2)}\right) \in \mathscr{Q}$, let define, for each $q \in \Omega$,

$$e_{\mathsf{WF}_q} \triangleq \left\|\mathbf{WF}_q\left(\mathbf{Q}_{-q}^{(1)}\right) - \mathbf{WF}_q\left(\mathbf{Q}_{-q}^{(2)}\right)\right\|_F, \quad (28)$$

$$e_q \triangleq \left\|\mathbf{Q}_q^{(1)} - \mathbf{Q}_q^{(2)}\right\|_F, \quad (29)$$

where, according to Lemma 1, each component $\mathbf{WF}_q(\mathbf{Q}_{-q})$ of $\mathbf{WF}$ can be rewritten as in (12), with $\mathbf{R}_{-q}(\mathbf{Q}_{-q})$ defined in (4). Then, we have:

$$e_{\mathsf{WF}_q} = \left\|\left[-\left(\mathbf{H}_{qq}^H \mathbf{R}_q^{-1}(\mathbf{Q}_{-q}^{(1)}) \mathbf{H}_{qq}\right)^{\sharp} - c_q \mathbf{P}_{\mathcal{N}(\mathbf{H}_{qq})}\right]_{\mathscr{Q}_q} \\ - \left[-\left(\mathbf{H}_{qq}^H \mathbf{R}_q^{-1}(\mathbf{Q}_{-q}^{(2)}) \mathbf{H}_{qq}\right)^{\sharp} - c_q \mathbf{P}_{\mathcal{N}(\mathbf{H}_{qq})}\right]_{\mathscr{Q}_q}\right\|_F \quad (30)$$

$$\leq \left\|\left(\mathbf{H}_{qq}^H \mathbf{R}_q^{-1}(\mathbf{Q}_{-q}^{(1)}) \mathbf{H}_{qq}\right)^{\sharp} - \left(\mathbf{H}_{qq}^H \mathbf{R}_q^{-1}(\mathbf{Q}_{-q}^{(2)}) \mathbf{H}_{qq}\right)^{\sharp}\right\|_F \quad (31)$$

$$= \left\|\mathbf{H}_{qq}^{\sharp} \left(\sum_{r \neq q} \mathbf{H}_{rq}\left(\mathbf{Q}_r^{(1)} - \mathbf{Q}_r^{(2)}\right) \mathbf{H}_{rq}^H\right) \mathbf{H}_{qq}^{\sharp H}\right\|_F \quad (32)$$

$$\leq \sum_{r \neq q} \rho\left(\mathbf{H}_{rq}^H \mathbf{H}_{qq}^{\sharp H} \mathbf{H}_{qq}^{\sharp} \mathbf{H}_{rq}\right) \left\|\mathbf{Q}_r^{(1)} - \mathbf{Q}_r^{(2)}\right\|_F \quad (33)$$

$$\triangleq \sum_{r \neq q} [\mathbf{S}]_{qr} \left\|\mathbf{Q}_r^{(1)} - \mathbf{Q}_r^{(2)}\right\|_F = \sum_{r \neq q} [\mathbf{S}]_{qr} e_r, \quad (34)$$

$\forall \mathbf{Q}^{(1)}, \mathbf{Q}^{(2)} \in \mathscr{Q}$ and $\forall q \in \Omega$, where: (30) follows from (12) (Lemma 1) with $c_q \geq \max\left(c_q(\mathbf{Q}_{-q}^{(1)}), c_q(\mathbf{Q}_{-q}^{(2)})\right)$; (31) follows from the non-expansive property of the projector in the Frobenius norm as given in (13) (Lemma 2); (32) follows from (23) (Lemma 3) and the assumption $\text{rank}(\mathbf{H}_{qq}) = n_{R_q}$, $\forall q \in \Omega$; (33) follows from the triangle inequality [26] and (25) (Lemma 4); and in (34) we used the definition of $\mathbf{S}$ as given in (20).

Introducing the vectors

$$\mathbf{e}_{\mathsf{WF}} \triangleq [e_{\mathsf{WF}_1}, \ldots, e_{\mathsf{WF}_Q}]^T, \quad \text{and} \quad \mathbf{e} \triangleq [e_1, \ldots, e_Q]^T, \quad (35)$$

with $e_{\mathsf{WF}_q}$ and $e_q$ defined in (28) and (29), respectively, the set of inequalities in (34) can be rewritten in vector form as

$$0 \leq \mathbf{e}_{\mathsf{WF}} \leq \mathbf{S}\,\mathbf{e}, \quad \forall \mathbf{Q}^{(1)},\, \mathbf{Q}^{(2)} \in \mathscr{Q}. \tag{36}$$

Using the weighted maximum norm $\|\cdot\|_{\infty,\mathrm{vec}}^{\mathbf{w}}$ defined in (17) in combination with (36), we have, for any given $\mathbf{w} > \mathbf{0}$,

$$\|\mathbf{e}_{\mathsf{WF}}\|_{\infty,\mathrm{vec}}^{\mathbf{w}} \leq \|\mathbf{S}\mathbf{e}\|_{\infty,\mathrm{vec}}^{\mathbf{w}} \leq \|\mathbf{S}\|_{\infty,\mathrm{mat}}^{\mathbf{w}} \|\mathbf{e}\|_{\infty,\mathrm{vec}}^{\mathbf{w}}, \tag{37}$$

$\forall \mathbf{Q}^{(1)}, \mathbf{Q}^{(2)} \in \mathscr{Q}$, where $\|\cdot\|_{\infty,\mathrm{mat}}^{\mathbf{w}}$ is the matrix norm induced by the vector norm $\|\cdot\|_{\infty,\mathrm{vec}}^{\mathbf{w}}$ in (17) and defined in (18) [26]. Finally, using (37) and (16), we obtain,

$$\begin{aligned}\left\|\mathbf{WF}(\mathbf{Q}^{(1)}) - \mathbf{WF}(\mathbf{Q}^{(2)})\right\|_{F,\mathrm{block}}^{\mathbf{w}} &= \|\mathbf{e}_{\mathsf{WF}}\|_{\infty,\mathrm{vec}}^{\mathbf{w}} \\ &\leq \|\mathbf{S}\|_{\infty,\mathrm{mat}}^{\mathbf{w}} \left\|\mathbf{Q}^{(1)} - \mathbf{Q}^{(2)}\right\|_{F,\mathrm{block}}^{\mathbf{w}},\end{aligned} \tag{38}$$

$\forall \mathbf{Q}^{(1)}, \mathbf{Q}^{(2)} \in \mathscr{Q}$ and $\forall \mathbf{w} > \mathbf{0}$, which leads to a block-contraction for the mapping $\mathbf{WF}$ if $\|\mathbf{S}\|_{\infty,\mathrm{mat}}^{\mathbf{w}} < 1$, implying condition (27). ∎

*Negative result #3:* The waterfilling mapping $\mathbf{WF}$ satisfies the Lipschitz property in (26) if the channel matrices $\{\mathbf{H}_{qq}\}_{q \in \Omega}$ are full row-rank. Surprisingly, if the channels are not full row-rank matrices, the property in (26) *does not* hold for *every* given set of matrices $\{\mathbf{H}_{qq}\}_{q \in \Omega}$, implying that the $\mathbf{WF}$ mapping is not a contraction under (27).

*Numerical counter-example*: Consider a game with two players. The Lipschitz property of the $\mathbf{WF}$ as given in Theorem 5 can be written, e.g., for player 1 as (a similar expression has to be satisfied for player 2)

$$\left\|\mathbf{WF}_1(\mathbf{Q}_2^{(1)}) - \mathbf{WF}_1(\mathbf{Q}_2^{(2)})\right\|_F \leq \rho\left(\mathbf{H}_{21}^H \mathbf{H}_{11}^{\sharp H} \mathbf{H}_{11}^{\sharp} \mathbf{H}_{21}\right) \\ \times \left\|\mathbf{Q}_2^{(1)} - \mathbf{Q}_2^{(2)}\right\|_F, \tag{39}$$

$\forall \mathbf{Q}^{(1)}, \mathbf{Q}^{(2)} \in \mathscr{Q}$. The following set of channels and users' covariance matrices does not satisfy the inequality in (39) [and thus (26)]: $P_1 = P_2 = 10$, $\mathbf{R}_1 = \mathbf{I}$, $\mathbf{R}_2 = \mathbf{I}$,

$$\mathbf{H}_{11} = \mathbf{H}_{22} = \begin{bmatrix} 0.5458 + 0.0819i & -0.5449 + 1.8701i \\ -2.1758 + 0.7811i & -1.9082 + 0.9013i \\ -1.0132 - 1.1376i & -1.8198 - 0.1200i \end{bmatrix}, \tag{40}$$

$$\mathbf{H}_{21} = \mathbf{H}_{12} = \begin{bmatrix} 0.5865 + 0.4392i & 1.4387 - 2.2133i \\ 1.5959 - 0.2853i & -1.5410 - 0.2285i \\ -0.1035 + 2.0967i & -0.3196 + 1.0228i \end{bmatrix} \tag{41}$$

and

$$\mathbf{Q}_2^{(1)} = \begin{bmatrix} 9.9175 & -0.8946 \\ -0.8946 & 0.0825 \end{bmatrix}, \tag{42}$$

$$\mathbf{Q}_2^{(2)} = \begin{bmatrix} 8.5842 & -1.2150 \\ -1.2150 & 1.4158 \end{bmatrix}. \tag{43}$$

The above setup leads to the following: $\left\|\mathbf{WF}_1(\mathbf{Q}_2^{(1)}) - \mathbf{WF}_1(\mathbf{Q}_2^{(2)})\right\|_F = 5.2925$, $\rho\left(\mathbf{H}_{21}^H \mathbf{H}_{11}^{\sharp H} \mathbf{H}_{11}^{\sharp} \mathbf{H}_{21}\right) = 2.5012$, and $\left\|\mathbf{Q}_2^{(1)} - \mathbf{Q}_2^{(2)}\right\|_F = 1.9392$. Since $\rho\left(\mathbf{H}_{21}^H \mathbf{H}_{11}^{\sharp H} \mathbf{H}_{11}^{\sharp} \mathbf{H}_{21}\right) \left\|\mathbf{Q}_2^{(1)} - \mathbf{Q}_2^{(2)}\right\|_F = $ 4.8502, condition (39) is not satisfied. Since this happens for both users, there exists no vector $\mathbf{w} > \mathbf{0}$ such that condition (26) can be satisfied [observe that for the setup above we have $\|\mathbf{S}\|_{\infty,\mathrm{mat}}^{\mathbf{w}} = \rho(\mathbf{H}_{21}^H \mathbf{H}_{11}^{\sharp H} \mathbf{H}_{11}^{\sharp} \mathbf{H}_{21})$].

The above example shows that (26) is not true for (strictly) tall matrices and, therefore, condition (27) is not enough to guarantee the $\mathbf{WF}$ mapping to be a block-contraction, implying that in such a case stronger conditions are needed. This is due to the fact that the reverse order law for generalized inverses of product of matrices (Lemma 3) does not hold if the outer matrices involved in the product are (strictly) tall [recall that the *inequality* in (24) always holds]. It turns out that, in such a case, a different approach is required to derive conditions guaranteeing the $\mathbf{WF}$ mapping to be a contraction. We address this issue in the next section.

*3) Case of full column-rank channel matrices:* The main difficulty in deriving contraction properties of the MIMO multiuser waterfilling mapping in the case of (strictly) tall channel matrices $\{\mathbf{H}_{qq}\}_{q \in \Omega}$ is that one cannot use the reverse order law of generalized inverses given in Lemma 3, as done in the proof of Theorem 5 [see (31)-(32)]. To overcome this issue, we develop a different approach based on the mean-value theorem for complex matrix-valued functions, as detailed next.

*Mean-value theorem for complex matrix-valued functions*: The mean value theorem for scalar real functions is one of the most important and basic theorems in functional analysis (cf. [34, Th.5.10], [31, Ch.5-Th.10]). In this paper, we will use the simplest form of the theorem, as stated next.

*Mean value theorem for real scalar functions* [34, Th.5.10]: Let $f : [a,b] \mapsto \mathbb{R}$ be a real continuous function on $[a,b]$, differentiable on $(a,b)$ with (first) derivative denoted by $f'$. Then,

$$\exists\, t \in (0,1)\,|\, f(b) - f(a) = f'(t\,b + (1-t)\,a)(b-a). \tag{44}$$

*Negative result #4*: Unfortunately, the generalization of (44) to vector-valued real functions that one would expect does not hold, meaning that for real vector-valued functions $\mathbf{f} : \mathcal{D} \subseteq \mathbb{R}^m \mapsto \mathbb{R}^n$ in general

$$\nexists\, t \in (0,1)\,|\, \mathbf{f}(\mathbf{y}) - \mathbf{f}(\mathbf{x}) = \mathbf{D}_{\mathbf{x}}\mathbf{f}(t\,\mathbf{y} + (1-t)\,\mathbf{x})(\mathbf{y} - \mathbf{x}), \tag{45}$$

for any $\mathbf{x}, \mathbf{y} \in \mathcal{D}$ and $\mathbf{x} \neq \mathbf{y}$, where $\mathbf{D}_{\mathbf{x}}\mathbf{f}$ denotes the Jacobian matrix of $\mathbf{f}$ (cf. Appendix B). One of the simplest examples to illustrate (45) is the following. Consider the real vector-valued function $\mathbf{f}(x) = [x^\alpha, x^\beta]^T$, with $x \in \mathbb{R}$ and, e.g., $\alpha = 2$, $\beta = 3$. There exists no value of $t \in (0,1)$ such that $\mathbf{f}(1) = \mathbf{f}(0) + \mathbf{D}_t \mathbf{f}(t)$.

Many extensions and variations of the main value theorem exist in the literature, either for (real/ complex) scalar or real vector-valued functions (see, e.g., [35], [36]). Here, we provide an extension of (44) in a form that is useful to our purpose, valid for complex matrix-valued functions. Interestingly, our result shows that (44) can be generalized with inequality to (complex) vector/matrix-valued functions if one: i) takes the norm on both sides; and ii) relaxes the equality with the

inequality.[6]

*Lemma 6:* Let $\mathbf{F}(\mathbf{X}) : \mathcal{D} \subseteq \mathbb{C}^{m \times n} \mapsto \mathbb{C}^{p \times q}$ *be a complex matrix-valued function defined on a convex set* $\mathcal{D}$, *assumed to be continuous on* $\mathcal{D}$ *and differentiable on the interior of* $\mathcal{D}$, *with Jacobian matrix* $\mathbf{D}_{\mathbf{X}}\mathbf{F}(\mathbf{X})$.[7] *Then, for any given* $\mathbf{X}, \mathbf{Y} \in \mathcal{D}$, *there exists some* $t \in (0,1)$ *such that*

$$\|\mathbf{F}(\mathbf{Y}) - \mathbf{F}(\mathbf{X})\|_F \leq \|\mathbf{D}_{\mathbf{X}}\mathbf{F}((t\,\mathbf{Y} + (1-t)\,\mathbf{X}))\operatorname{vec}(\mathbf{Y} - \mathbf{X})\|_2 \quad (46)$$

$$\leq \|\mathbf{D}_{\mathbf{X}}\mathbf{F}\left((t\,\mathbf{Y} + (1-t)\,\mathbf{X})\right)\|_{2,\mathrm{mat}} \|\mathbf{Y} - \mathbf{X}\|_F, \quad (47)$$

*where* $\|\mathbf{A}\|_{2,\mathrm{mat}} \triangleq \sqrt{\rho(\mathbf{A}^H \mathbf{A})}$ *denotes the spectral norm of* $\mathbf{A}$.

*Proof:* See Appendix B. ∎

We can now provide the contraction theorem for the **WF** mapping valid also for the case in which the channels $\{\mathbf{H}_{qq}\}_{q\in\Omega}$ are full column-rank matrices.

*Theorem 7 (Contraction property of* **WF** *mapping):* Suppose that $\operatorname{rank}(\mathbf{H}_{qq}) = n_{T_q}, \forall q \in \Omega$ (*i.e., full rank tall/square matrices*). *Then, for any given* $\mathbf{w} \triangleq [w_1, \ldots, w_Q]^T > \mathbf{0}$, *the mapping* **WF** *defined in* (15) *is Lipschitz continuous on* $\mathcal{Q}$:

$$\left\|\mathbf{WF}(\mathbf{Q}^{(1)}) - \mathbf{WF}(\mathbf{Q}^{(2)})\right\|_{F,\mathrm{block}}^{\mathbf{w}} \leq \|\mathbf{S}(\mathscr{P}^\star)\|_{\infty,\mathrm{mat}}^{\mathbf{w}}$$
$$\times \left\|\mathbf{Q}^{(1)} - \mathbf{Q}^{(2)}\right\|_{F,\mathrm{block}}^{\mathbf{w}}, \quad (48)$$

*with* $\|\cdot\|_{F,\mathrm{block}}^{\mathbf{w}}$, $\|\cdot\|_{\infty,\mathrm{mat}}^{\mathbf{w}}$ *and* $\mathbf{S}(\mathscr{P}^\star)$ *defined in* (16), (18) *and* (19), *respectively, where* $\mathscr{P}^\star = (\mathbf{P}_{rq}^\star)_{r\neq q}$ *and each* $\mathbf{P}_{rq}^\star$ *is a solution to the following optimization problem*:

$$\begin{array}{ll}\underset{\mathbf{P}_{rq},\mathbf{Q}_{-q}\in\mathscr{Q}_{-q}}{\text{maximize}} & \rho\left(\mathbf{H}_{rq}^H \mathbf{P}_{rq}^H \mathbf{H}_{qq}^{\sharp H} \mathbf{H}_{qq}^\sharp \mathbf{P}_{rq} \mathbf{H}_{rq}\right) \\ \text{subject to} & \mathbf{P}_{rq} = \mathbf{P}_{rq}^2 \\ & \mathcal{R}(\mathbf{P}_{rq}) = \mathcal{R}(\mathbf{H}_{qq}) \\ & \mathcal{N}(\mathbf{P}_{rq}) = \mathcal{N}(\mathbf{H}_{qq}^H \mathbf{R}_{-q}(\mathbf{Q}_{-q}))\end{array} \quad (49)$$

*Furthermore, if the following condition is satisfied*

$$\|\mathbf{S}(\mathscr{P}^\star)\|_{\infty,\mathrm{mat}}^{\mathbf{w}} < 1, \qquad \text{for some } \mathbf{w} > \mathbf{0}, \quad (50)$$

*then, the mapping* **WF** *is a* block-contraction *with modulus* $\beta = \|\mathbf{S}(\mathscr{P}^\star)\|_{\infty,\mathrm{mat}}^{\mathbf{w}}$. ∎

Observe that stronger sufficient conditions for the **WF** mapping to be a contraction than (50) can be obtained by solving any relaxed (simpler) version of the optimization problem in (49). For example, to simplify the optimization, one can remove the last constraint in (49). However, solving (a relaxation of) (49) may still no be easy. To overcome this issue and give additional insight into the physical interpretation of the conditions we obtained, we provide the following corollary, which contains sufficient conditions for (50) that are easier to be checked in practice, since they do not depend on the set $\mathscr{P}^\star$.

*Corollary 8:* Given $\mathbf{S}(\mathscr{P}^\star)$ and $\mathbf{S}^{\mathrm{up}}$ defined in (19) and (22), respectively, with $\mathscr{P}^\star$ given in Theorem 7, we have:

$$\mathbf{S}(\mathscr{P}^\star) < \mathbf{S}^{\mathrm{up}}, \quad (51)$$

*implying*

$$\|\mathbf{S}(\mathscr{P}^\star)\|_{\infty,\mathrm{mat}}^{\mathbf{w}} < \|\mathbf{S}^{\mathrm{up}}\|_{\infty,\mathrm{mat}}^{\mathbf{w}}, \qquad \text{for any } \mathbf{w} > \mathbf{0}. \quad (52)$$

∎

*Proof of Theorem 7.* The proof follows the same guidelines of that of Theorem 5, with the key difference that, in the case of (strictly) full column-rank channel matrices, we cannot use the reverse order law (Lemma 3) as done to obtain (31)-(32) in the proof of Theorem 5. We apply instead the mean-value theorem in Lemma 6, as detailed next. For technical reasons, we introduce first a proper complex matrix-valued function $\mathbf{F}_q(\mathbf{Q}_{-q})$ related to the MIMO multiuser waterfilling mapping $\mathbf{WF}_q(\mathbf{Q}_{-q})$ in (9) and, using Lemma 6, we study the Lipschitz properties of the function on $\mathscr{Q}_{-q}$. Then, building on this result, we show that the **WF** mapping satisfies (48).

Given $q \in \Omega$, let us introduce the following complex matrix-valued function $\mathbf{F}_q(\mathbf{Q}_{-q}) : \mathscr{Q}_{-q} \mapsto \mathbb{S}_{++}^{n_{T_q} \times n_{T_q}}$, defined as:

$$\mathbf{F}_q(\mathbf{Q}_{-q}) = \left(\mathbf{H}_{qq}^H \mathbf{R}_{-q}^{-1}(\mathbf{Q}_{-q}) \mathbf{H}_{qq}\right)^{-1}, \quad (53)$$

with $\mathbf{R}_{-q}(\mathbf{Q}_{-q})$ given in (4). Observe that the function $\mathbf{F}_q(\mathbf{Q}_{-q})$ is continuous on $\mathscr{Q}_{-q}$ (implied from the continuity of $\mathbf{R}_{-q}^{-1}(\mathbf{Q}_{-q})$ at any $\mathbf{Q}_{-q} \succeq \mathbf{0}$[8]) and differentiable at any $\mathbf{Q}_{-q} \succeq \mathbf{0}$ (cf. Appendix B). The Jacobian matrix of $\mathbf{F}_q(\mathbf{Q}_{-q})$ with respect to $\mathbf{Q}_{-q}$ is (see Lemma 17 in Appendix B):

$$\begin{aligned}\mathbf{D}_{\mathbf{Q}_{-q}}\mathbf{F}(\mathbf{Q}_{-q}) = & \left[\mathbf{G}_{1\,q}^*(\mathbf{Q}_{-q}) \otimes \mathbf{G}_{1\,q}(\mathbf{Q}_{-q}), \ldots,\right.\\ & \mathbf{G}_{q-1\,q}^*(\mathbf{Q}_{-q}) \otimes \mathbf{G}_{q-1\,q}(\mathbf{Q}_{-q}), \ldots,\\ & \mathbf{G}_{q+1\,q}^*(\mathbf{Q}_{-q}) \otimes \mathbf{G}_{q+1\,q}(\mathbf{Q}_{-q}), \ldots,\\ & \left.\mathbf{G}_{Q\,q}^*(\mathbf{Q}_{-q}) \otimes \mathbf{G}_{Q\,q}(\mathbf{Q}_{-q})\right],\end{aligned} \quad (54)$$

where

$$\mathbf{G}_{rq}(\mathbf{Q}_{-q}) \triangleq \left(\mathbf{H}_{qq}^H \mathbf{R}_{-q}^{-1}(\mathbf{Q}_{-q}) \mathbf{H}_{qq}\right)^{-1} \mathbf{H}_{qq}^H \mathbf{R}_{-q}^{-1}(\mathbf{Q}_{-q}) \mathbf{H}_{rq}. \quad (55)$$

Observe that $\mathbf{D}_{\mathbf{Q}_{-q}}\mathbf{F}(\mathbf{Q}_{-q})$ is well-defined and continuous at any $\mathbf{Q}_{-q} \succeq \mathbf{0}$.

Therefore, function $\mathbf{F}_q(\mathbf{Q}_{-q})$ satisfies the assumption of the mean-value theorem in Lemma 6, meaning that, for any two different points $\mathbf{Q}_{-q}^{(1)}, \mathbf{Q}_{-q}^{(2)} \in \mathscr{Q}_{-q}$, with $\mathbf{Q}_{-q}^{(i)} = [\mathbf{Q}_1^{(i)}, \ldots, \mathbf{Q}_{q-1}^{(i)}, \mathbf{Q}_{q+1}^{(i)}, \ldots, \mathbf{Q}_Q^{(i)}]$ for $i = 1, 2$, there exists some $t \in (0,1)$ such that, introducing

$$\mathbf{\Delta} \triangleq t\mathbf{Q}_{-q}^{(1)} + (1-t)\mathbf{Q}_{-q}^{(2)}, \quad (56)$$

---

[6]Basic results on complex differential calculus for matrix-valued functions along with some intermediate results useful to prove Lemma 6 are given in Appendix B.

[7]Since the complex matrix-valued functions we are interested in depend only on $\mathbf{X}$ (and not on $\mathbf{X}^*$), we consider the mean value theorem only for this case. The generalization to the complex matrix-valued functions depending on both $\mathbf{X}$ and $\mathbf{X}^*$ follows similar steps. Furthermore, since our functions are continuous and differentiable on a convex set, we made these assumptions in the theorem. However, the same statement can be established under weaker assumptions.

[8]This result can be proved using [27, Th. 10.7.1].

we have:

$$\left\|\mathbf{F}_q\left(\mathbf{Q}_{-q}^{(1)}\right) - \mathbf{F}_q\left(\mathbf{Q}_{-q}^{(2)}\right)\right\|_F$$

$$\leq \left\|\mathbf{D}_{\mathbf{Q}_{-q}}\mathbf{F}_q(\mathbf{\Delta})\,\text{vec}\left(\mathbf{Q}_{-q}^{(1)} - \mathbf{Q}_{-q}^{(2)}\right)\right\|_2 \quad (57)$$

$$\leq \sum_{r \neq q}\left\|\mathbf{G}_{rq}^*(\mathbf{\Delta}) \otimes \mathbf{G}_{rq}(\mathbf{\Delta})\right\|_{2,\text{mat}}\left\|\mathbf{Q}_r^{(1)} - \mathbf{Q}_r^{(2)}\right\|_F \quad (58)$$

$$= \sum_{r \neq q}\rho\left(\mathbf{G}_{rq}^H(\mathbf{\Delta})\mathbf{G}_{rq}(\mathbf{\Delta})\right)\left\|\mathbf{Q}_r^{(1)} - \mathbf{Q}_r^{(2)}\right\|_F, \quad (59)$$

where (57) follows from (46) (Lemma 6); (58) follows from the structure of $\mathbf{D}_{\mathbf{Q}_{-q}}\mathbf{F}_q$ [see (54)] and the triangle inequality [26]; and (59) comes from the following chain of equalities:

$$\rho\left[\left(\mathbf{G}_{rq}^T \otimes \mathbf{G}_{rq}^H\right)\left(\mathbf{G}_{rq}^* \otimes \mathbf{G}_{rq}\right)\right] = \rho\left[\mathbf{G}_{rq}^T\mathbf{G}_{rq}^* \otimes \mathbf{G}_{rq}^H\mathbf{G}_{rq}\right]$$
$$= \left(\rho\left[\mathbf{G}_{rq}^H\mathbf{G}_{rq}\right]\right)^2, \quad (60)$$

where the last equality in (60) follows from [29, Th.21.11.4] [implying $\rho(\mathbf{G}_{rq}^T\mathbf{G}_{rq}^* \otimes \mathbf{G}_{rq}^H\mathbf{G}_{rq}) = \rho(\mathbf{G}_{rq}^T\mathbf{G}_{rq}^*)\,\rho(\mathbf{G}_{rq}^H\mathbf{G}_{rq})$] and the fact that the eigenvalues of $\mathbf{G}_{rq}^T\mathbf{G}_{rq}^*$ coincide with those of $\mathbf{G}^H\mathbf{G}_{rq}$.

Observe that, differently from (33)-(34), the factor $\alpha_{rq}(\mathbf{\Delta}) \triangleq \rho\left[\mathbf{G}_{rq}^H(\mathbf{\Delta})\mathbf{G}_{rq}(\mathbf{\Delta})\right]$ in (59) depends, in general, on both $t \in (0,1)$ and the covariance matrices $\mathbf{Q}_{-q}^{(1)}$ and $\mathbf{Q}_{-q}^{(2)}$ through $\mathbf{\Delta}$ [see (56)]:

$$\alpha_{rq}(\mathbf{\Delta}) = \rho\left[\mathbf{H}_{rq}^H\mathbf{R}_{-q}^{-1}(\mathbf{\Delta})\mathbf{H}_{qq}\left(\mathbf{H}_{qq}^H\mathbf{R}_{-q}^{-1}(\mathbf{\Delta})\mathbf{H}_{qq}\right)^{-1}\right.$$
$$\left.\times \left(\mathbf{H}_{qq}^H\mathbf{R}_{-q}^{-1}(\mathbf{\Delta})\mathbf{H}_{qq}\right)^{-1}\mathbf{H}_{qq}^H\mathbf{R}_{-q}^{-1}(\mathbf{\Delta})\mathbf{H}_{rq}\right] \quad (61)$$

where in (61) we used (55). Interestingly, in the case of square (nonsingular) channel matrices $\mathbf{H}_{qq}$, (61) reduces to $\alpha_{rq}(\mathbf{\Delta}) = \rho\left[\mathbf{H}_{rq}^H\mathbf{H}_{qq}^{\sharp H}\mathbf{H}_{qq}^{\sharp}\mathbf{H}_{rq}\right] = [\mathbf{S}]_{qr}$, where $\mathbf{S}$ is defined in (20), which leads to the same contraction factor for the $\mathbf{WF}$ mapping as in Theorem 5.

We thus focus on the case of (strictly) full column-rank matrices $\mathbf{H}_{qq}$ and look for an upper bound of $\alpha_{rq}(\mathbf{\Delta})$, independent of $\mathbf{\Delta}$. To this end, we introduce the following: the orthogonal projection onto $\mathcal{R}\left(\mathbf{R}_{-q}^{-1/2}(\mathbf{\Delta})\mathbf{H}_{qq}\right)$, given by

$$\mathbf{P}_{\mathcal{R}\left(\mathbf{R}_{-q}^{-1/2}(\mathbf{\Delta})\mathbf{H}_{qq}\right)} = $$
$$\mathbf{R}_{-q}^{-1/2}(\mathbf{\Delta})\mathbf{H}_{qq}\left(\mathbf{H}_{qq}^H\mathbf{R}_{-q}^{-1}(\mathbf{\Delta})\mathbf{H}_{qq}\right)^{-1}\mathbf{H}_{qq}^H\mathbf{R}_{-q}^{-1/2}(\mathbf{\Delta}) \quad (62)$$

and the idempotent matrix $\mathbf{P}_{rq}(\mathbf{\Delta})$ [i.e., $\mathbf{P}_{rq}(\mathbf{\Delta}) = \mathbf{P}_{rq}^2(\mathbf{\Delta})$], defined as

$$\mathbf{P}_{rq}(\mathbf{\Delta}) \triangleq \mathbf{R}_{-q}^{1/2}(\mathbf{\Delta})\mathbf{P}_{\mathcal{R}\left(\mathbf{R}_{-q}^{-1/2}(\mathbf{\Delta})\mathbf{H}_{qq}\right)}\mathbf{R}_{-q}^{-1/2}(\mathbf{\Delta}). \quad (63)$$

Observe that $\mathbf{P}_{rq}(\mathbf{\Delta})$ is still a projection (albeit non orthogonal): the projection onto the subspace $\mathcal{U}$ of $\mathbb{C}^{n_{R_q}}$ along the subspace $\mathcal{V}$, where $\mathcal{U}$ and $\mathcal{V}$ are given by (observe that $\mathbf{H}_{qq}$ has an empty null space):

$$\begin{aligned}\mathcal{U} &= \mathcal{R}\left(\mathbf{P}_{rq}(\mathbf{\Delta})\right) = \mathcal{R}(\mathbf{H}_{qq}),\\\mathcal{V} &= \mathcal{N}\left(\mathbf{P}_{rq}(\mathbf{\Delta})\right) = \mathcal{N}\left(\mathbf{H}_{qq}^H\mathbf{R}_{-q}^{-1}(\mathbf{\Delta})\right),\end{aligned} \quad (64)$$

and $\mathcal{U} \oplus \mathcal{V} = \mathbb{C}^{n_{R_q}}$. Using the above definitions, we can rewrite (61) as:

$$\alpha_{rq}(\mathbf{\Delta}) = \rho\left[\mathbf{H}_{rq}^H\mathbf{P}_{rq}^H(\mathbf{\Delta})\mathbf{H}_{qq}^{\sharp H}\mathbf{H}_{qq}^{\sharp}\mathbf{P}_{rq}(\mathbf{\Delta})\mathbf{H}_{rq}\right] \quad (65)$$

$$\leq \rho\left[\mathbf{H}_{rq}^H\mathbf{P}_{rq}^{\star H}\mathbf{H}_{qq}^{\sharp H}\mathbf{H}_{qq}^{\sharp}\mathbf{P}_{rq}^{\star}\mathbf{H}_{rq}\right] \triangleq [\mathbf{S}(\mathscr{P}^{\star})]_{qr}, \quad (66)$$

where in (65) we used (62) and (63); and (66) follows from the definition of $\mathscr{P}^{\star}$ as given in (49) and the characterization of $\mathbf{P}_{rq}(\mathbf{\Delta})$ as given on (64).

The Lipschitz property of the $\mathbf{WF}$ mapping as given in (48) comes from (59) and (66), using the same steps as in the proof of Theorem 5. We omit the details, because of space limitations.

*Proof of Corollary 8.* We prove now the corollary, providing an upper bound of each $[\mathbf{S}(\mathscr{P}^{\star})]_{qr}$ in (66), independent of $\mathbf{P}_{rq}$. For any fixed $q$ and $r \neq q$, given $\mathbf{\Delta}$ in (56), it follows from (65) that

$$\alpha_{rq}(\mathbf{\Delta}) \leq \rho\left(\mathbf{H}_{rq}^H\mathbf{H}_{rq}\right)\rho\left(\mathbf{H}_{qq}^{\sharp H}\mathbf{H}_{qq}^{\sharp}\right)$$
$$\times \rho\left(\mathbf{R}_{-q}^{-1/2}(\mathbf{\Delta})\mathbf{P}_{\mathcal{R}\left(\mathbf{R}_{-q}^{-1/2}(\mathbf{\Delta})\mathbf{H}_{qq}\right)}\mathbf{R}_{-q}(\mathbf{\Delta})\right. \quad (67)$$
$$\left.\times \mathbf{P}_{\mathcal{R}\left(\mathbf{R}_{-q}^{-1/2}(\mathbf{\Delta})\mathbf{H}_{qq}\right)}\mathbf{R}_{-q}^{-1/2}(\mathbf{\Delta})\right) \quad (68)$$
$$\leq \rho\left(\mathbf{H}_{rq}^H\mathbf{H}_{rq}\right)\rho\left(\mathbf{H}_{qq}^{\sharp H}\mathbf{H}_{qq}^{\sharp}\right)\rho\left(\mathbf{R}_{-q}(\mathbf{\Delta})\right)\rho\left(\mathbf{R}_{-q}^{-1}(\mathbf{\Delta})\right)$$
$$\quad (69)$$
$$< \rho\left(\mathbf{H}_{rq}^H\mathbf{H}_{rq}\right)\rho\left(\mathbf{H}_{qq}^{\sharp H}\mathbf{H}_{qq}^{\sharp}\right)\text{innr}_q \quad (70)$$

where (68)-(69) follow from

$$\rho(\mathbf{A}^H\mathbf{B}\mathbf{A}) \leq \rho(\mathbf{A}^H\mathbf{C}\mathbf{A}), \quad \text{for all} \quad \mathbf{0} \preceq \mathbf{B} \preceq \mathbf{C}, \quad (71)$$

and the fact that $\mathbf{P}_{\mathcal{R}\left(\mathbf{R}_{-q}^{-1/2}(\mathbf{\Delta}_0)\mathbf{H}_{qq}\right)}$ is an orthogonal projection, implying $\mathbf{P}_{\mathcal{R}\left(\mathbf{R}_{-q}^{-1/2}(\mathbf{\Delta}_0)\mathbf{H}_{qq}\right)}^2 = \mathbf{P}_{\mathcal{R}\left(\mathbf{R}_{-q}^{-1/2}(\mathbf{\Delta}_0)\mathbf{H}_{qq}\right)}$ and $\mathbf{P}_{\mathcal{R}\left(\mathbf{R}_{-q}^{-1/2}(\mathbf{\Delta}_0)\mathbf{H}_{qq}\right)} \preceq \mathbf{I}$; and in (70) we used $\rho\left[\mathbf{R}_{-q}(\mathbf{\Delta})\right]\rho\left[\mathbf{R}_{-q}^{-1}(\mathbf{\Delta})\right] < \text{innr}_q$, where $\text{innr}_q$ is defined in (21) and the upper-bound comes from the following chain of inequalities:

$$\mathbf{R}_{n_q} \preceq \mathbf{R}_{-q}(\mathbf{\Delta}) = \mathbf{R}_{n_q} + \sum_{r \neq q}\mathbf{H}_{rq}\left(t\mathbf{Q}_r^{(1)} + (1-t)\mathbf{Q}_r^{(2)}\right)\mathbf{H}_{rq}^H$$
$$\preceq \mathbf{R}_{n_q} + \sum_{r \neq q}\text{Tr}\left(t\mathbf{Q}_r^{(1)} + (1-t)\mathbf{Q}_r^{(2)}\right)\mathbf{H}_{rq}\mathbf{H}_{rq}^H$$
$$= \mathbf{R}_{n_q} + \sum_{r \neq q}P_r\mathbf{H}_{rq}\mathbf{H}_{rq}^H$$
$$\quad (72)$$

where we used the fact that $t\mathbf{Q}_r^{(1)} + (1-t)\mathbf{Q}_r^{(2)} \succeq \mathbf{0}$, for all $t \in (0,1)$ and $r \in \Omega$, and $\text{Tr}(\mathbf{Q}_r^{(1)}) = \text{Tr}(\mathbf{Q}_r^{(2)}) = P_r$ for all $r \in \Omega$.

The inequality in (51) follows directly from (66) and (70) [observe that the upper bound of $\alpha_{rq}(\mathbf{\Delta})$ in (70) is strict and does not depend on $\mathbf{\Delta}$], which completes the proof. ∎

Comparing Theorems 5 and 7 (see also Corollary 8), one infers that conditions for the multiuser MIMO waterfilling mapping to be a block-contraction in the case of full column-rank (rectangular) channel matrices are stronger than those required when the channels are full row-rank (rectangular) matrices. Interestingly, both conditions coincide if the (direct) channel matrices are square nonsingular [in fact, $\text{rank}(\mathbf{P}_{rq}) = n_{R_q}$, implying that the feasible set of (49) contains only the point $\mathbf{P}_{rq} = \mathbf{I}$]. More specifically, for any given set of channel matrices (either tall or fat), the relationship existing among conditions in Theorems 5 and 7 is the following:

$$\mathbf{S} \leq \mathbf{S}(\mathscr{P}^\star) < \mathbf{S}^{\text{up}}, \tag{73}$$

where $\mathbf{S}$, $\mathbf{S}(\mathscr{P}^\star)$ and $\mathbf{S}^{\text{up}}$ are defined in (20), (19), and (22), respectively, with $\mathscr{P}^\star$ given in Theorem 7. The lower-bound in (73) is indeed reached when the channels $\{\mathbf{H}_{qq}\}_{q \in \Omega}$ are square nonsingular matrices. A physical interpretation of sufficient conditions (27) and (50) is given in the next section where, building on Theorems 5 and 7, we provide a unified set of sufficient conditions guaranteeing the uniqueness of the NE of game $\mathscr{G}$.

## IV. EXISTENCE AND UNIQUENESS OF THE NE

Using the results obtained in the previous section, we can now study game $\mathscr{G}$ and derive conditions for the uniqueness of the NE, as given next. Differently from current works in the literature [6]-[16], the proposed conditions are valid for *any given set of arbitrary channel matrices*, either tall/fat or singular.

Before stating the main theorem, we introduce the following intermediate definitions. Given $\mathbf{S}(\mathscr{P})$, $\mathbf{S}$ and $\mathbf{S}^{\text{up}}$ defined in (19), (20) and (22), respectively, let $\widetilde{\mathbf{S}} \in \mathbb{R}_+^{Q \times Q}$ and $\widetilde{\mathbf{S}}^{\text{up}} \in \mathbb{R}_+^{Q \times Q}$ be nonnegative matrices, defined, for each $r, q \in \Omega$, as:

$$\left[\widetilde{\mathbf{S}}\right]_{qr} \triangleq \begin{cases} [\mathbf{S}]_{qr}, & \text{if } \text{rank}(\mathbf{H}_{qq}) = n_{R_q}, \\ [\mathbf{S}(\mathscr{P}^\star)]_{qr}, & \text{otherwise}, \end{cases} \tag{74}$$

and

$$\left[\widetilde{\mathbf{S}}^{\text{up}}\right]_{qr} \triangleq \begin{cases} [\mathbf{S}]_{qr}, & \text{if } \text{rank}(\mathbf{H}_{qq}) = n_{R_q}, \\ [\mathbf{S}^{\text{up}}]_{qr}, & \text{otherwise}, \end{cases} \tag{75}$$

where $\mathscr{P}^\star$ is given in Theorem 7.

A unified set of sufficient conditions guaranteeing the uniqueness of the NE of $\mathscr{G}$ are given in the following theorem.

*Theorem 9:* Game $\mathscr{G}$ always admits a NE, for any set of channel matrices and transmit power of the users. Furthermore, the NE is unique if [9]

$$\rho(\widetilde{\mathbf{S}}) < 1, \tag{C1}$$

*where $\widetilde{\mathbf{S}}$ is defined in* (74).

*Proof:* See Appendix C. ∎

Even though condition (C1) is valid for any set of channel matrices, either tall/fat or singular, it may not be easy to check it, when some (direct) channel matrix is (strictly) tall or singular, because of the difficulty in the computation of the set $\mathscr{P}^\star$. To overcome this issue, we provide in the following a set of sufficient conditions for (C1) that are easier to be checked, still valid for arbitrary channel matrices.

[9] In the case of rank deficient channel matrices $\{\mathbf{H}_{qq}\}_{q \in \Omega}$, one can obtain a weaker condition than (C1), as given in Appendix C [see (127)].

*Corollary 10:* A sufficient condition for (C1) in Theorem 9 is

$$\rho(\widetilde{\mathbf{S}}^{\text{up}}) < 1, \tag{C2}$$

*where $\widetilde{\mathbf{S}}^{\text{up}}$ is defined in* (75). ∎

Finally, to give additional insight into the physical interpretation of sufficient conditions for the uniqueness of the NE, we provide the following.

*Corollary 11:* If $\text{rank}(\mathbf{H}_{qq}) = n_{R_q}$ for all $q \in \Omega$, then a sufficient condition for (C1) in Theorem 9 is given by one of the two following set of conditions:

Low MUI received:
$$\frac{1}{w_q} \sum_{r \neq q} \rho \left( \mathbf{H}_{rq}^H \mathbf{H}_{qq}^{\sharp H} \mathbf{H}_{qq}^{\sharp} \mathbf{H}_{rq} \right) w_r < 1, \quad \forall q \in \Omega, \tag{C3}$$

Low MUI generated:
$$\frac{1}{w_r} \sum_{q \neq r} \rho \left( \mathbf{H}_{rq}^H \mathbf{H}_{qq}^{\sharp H} \mathbf{H}_{qq}^{\sharp} \mathbf{H}_{rq} \right) w_q < 1, \quad \forall r \in \Omega, \tag{C4}$$

*where $\mathbf{w} \triangleq [w_1, \ldots, w_Q]^T$ is any positive vector.*

*If $\text{rank}(\mathbf{H}_{qq}) \leq n_{T_q}$, for all $q \in \Omega$, then a sufficient condition for (C1) is given by one of the two following set of conditions:*[10]

Low MUI received:
$$\frac{1}{w_q} \sum_{r \neq q} \text{innr}_q \cdot \rho \left( \mathbf{H}_{rq}^H \mathbf{H}_{rq} \right) \rho \left( \mathbf{H}_{qq}^{\sharp H} \mathbf{H}_{qq}^{\sharp} \right) w_r < 1, \quad \forall q \in \Omega, \tag{C5}$$

Low MUI generated:
$$\frac{1}{w_r} \sum_{q \neq r} \text{innr}_q \cdot \rho \left( \mathbf{H}_{rq}^H \mathbf{H}_{rq} \right) \rho \left( \mathbf{H}_{qq}^{\sharp H} \mathbf{H}_{qq}^{\sharp} \right) w_q < 1, \quad \forall r \in \Omega, \tag{C6}$$

*where the $\text{innr}_q$'s are defined in* (21). ∎

*Remark 3 - On the uniqueness conditions.* Conditions (C3)-(C4) and (C5)-(C6) provide a physical interpretation of the uniqueness of the NE: as expected, the uniqueness of the NE is ensured if the interference among the links is sufficiently small. The importance of (C3)-(C4) and (C5)-(C6) is that they quantify how small the interference must be to guarantee that the equilibrium is indeed unique. Specifically, conditions (C3) and (C5) can be interpreted as a constraint on the maximum amount of interference that each receiver can tolerate, whereas (C4) and (C6) introduce an upper bound on the maximum level of interference that each transmitter is allowed to generate. Surprisingly, the above conditions differ if the channel matrices $\{\mathbf{H}_{qq}\}_{q \in \Omega}$ are (strictly) tall or fat. More specifically, (C5)-(C6) are stronger than (C3)-(C4), implying that in the case of tall or rank deficient channel matrices a stronger constraint is imposed on the multiuser interference to guarantee the uniqueness of the NE. Furthermore, differently from (C3)-(C4), conditions (C5)-(C6) depend also on the transmit powers of the users and the noise power, through the $\mathrm{innr}_{rq}$'s coefficients. This difference in the uniqueness conditions is due to the fact that, in the case of (strictly) tall channel matrices $\{\mathbf{H}_{qq}\}_{q\in\Omega}$, the multiuser waterfilling requires a stronger condition to be a contraction (cf. Sec. III-B3).

[10] The case in which some channel matrices $\mathbf{H}_{qq}$ are (strictly) tall and some others are fat or there are rank deficient channel matrices can be similarly addressed: if $\text{rank}(\mathbf{H}_{qq}) < n_{R_q}$ for some $q$ (implying that $\mathbf{H}_{qq}$ is a (strictly) tall, possibly singular, or a fat singular matrix) we can still use (C3) [or (C4)] where the $q$-th condition in (C3) [or the $q$-th term in the sum in (C4)] is replaced by the $q$-th condition in (C5) [or the corresponding $q$-th term in the sum in (C6)].

Finally, observe that all the above conditions coincide in the case of square non singular matrices $\{\mathbf{H}_{qq}\}_{q\in\Omega}$ and, in such a case, they contain, as special case, most of the conditions known in the literature [6]-[12] for the rate-maximization game in SISO frequency-selective interference channels and OFDM transmission. A detailed comparison between the general conditions (C1) in the case of nonsingular square MIMO channels and those obtained in the cited papers was recently given in [16].

To obtain weaker sufficient conditions guaranteeing the uniqueness of the Nash equilibrium (and the convergence of distributed algorithms) even in the case of tall channel matrices, which coincide with those obtained in the full row-rank case (the weakest ones), in Sec. V-B, we propose a variation of the original game $\mathscr{G}$, whose performance in terms of sum-rate is virtually the same as that of game $\mathscr{G}$.

## V. ASYNCHRONOUS DISTRIBUTED ALGORITHMS

In this section we focus on distributed algorithms that converge to the NE of the game. We consider totally asynchronous distributed algorithms, meaning that in the updating procedure some users are allowed to change their strategy more frequently than the others, and they might even perform these updates using *outdated* information on the interference caused by the others. We propose two different classes of asynchronous algorithms and show that, whatever the asynchronous mechanism is, both algorithms converge to the globally asymptotically stable NE of the games with virtually the same performance, under conditions guaranteeing the uniqueness of the NE.

### A. The MIMO Iterative Waterfilling Algorithm

To provide a formal description of the proposed asynchronous MIMO IWFA, we briefly recall some intermediate definitions, as given in [16]. We assume, without loss of generality, that the set of times at which one or more users update their strategies is the discrete set $\mathcal{T} = \mathbb{N}_+ = \{0, 1, 2, \ldots\}$. Let $\mathbf{Q}_q^{(n)}$ denote the covariance matrix of the vector signal transmitted by user $q$ at the $n$-th iteration, and let $\mathcal{T}_q \subseteq \mathcal{T}$ denote the set of times $n$ at which $\mathbf{Q}_q^{(n)}$ is updated (thus, at time $n \notin \mathcal{T}_q$, $\mathbf{Q}_q^{(n)}$ is left unchanged). Let $\tau_r^q(n)$ denote the most recent time at which the interference from user $r$ is perceived by user $q$ at the $n$-th iteration (observe that $\tau_r^q(n)$ satisfies $0 \leq \tau_r^q(n) \leq n$). Hence, if user $q$ updates his own covariance matrix at the $n$-th iteration, then he chooses his optimal $\mathbf{Q}_q^{(n)}$, according to (9), and using the interference level caused by

$$\mathbf{Q}_{-q}^{(\boldsymbol{\tau}^q(n))} \triangleq \left(\mathbf{Q}_1^{(\tau_1^q(n))}, \ldots, \mathbf{Q}_{q-1}^{(\tau_{q-1}^q(n))}, \mathbf{Q}_{q+1}^{(\tau_{q+1}^q(n))}, \ldots, \mathbf{Q}_Q^{(\tau_Q^q(n))}\right). \quad (76)$$

Some standard conditions in asynchronous convergence theory that are fulfilled in any practical implementation need to be satisfied by the schedule $\{\tau_r^q(n)\}$ and $\{\mathcal{T}_q\}$; we refer to [16], [12] for the details. Using the above notation, the asynchronous MIMO IWFA is formally described in Algorithm 1. Sufficient conditions that guarantee the global convergence of the algorithm are given in Theorem 12.

---
**Algorithm 1: MIMO Asynchronous IWFA**

---
Set $n = 0$ and $\mathbf{Q}_q^{(0)} \in \mathscr{Q}_q$;
for $n = 0 : \mathrm{N}_{\mathrm{it}}$

$$\mathbf{Q}_q^{(n+1)} = \begin{cases} \mathbf{WF}_q\left(\mathbf{Q}_{-q}^{(\boldsymbol{\tau}^q(n))}\right), & \text{if } n \in \mathcal{T}_q, \\ \mathbf{Q}_q^{(n)}, & \text{otherwise;} \end{cases} \quad \forall q \in \Omega \quad (77)$$

end

---

*Theorem 12:* Assume that condition (C1) of Theorem 9 is satisfied. Then, as $\mathrm{N}_{\mathrm{it}} \to \infty$, the asynchronous MIMO IWFA, described in Algorithm 1, converges to the unique NE of game $\mathscr{G}$, for any set of feasible initial conditions and updating schedule.

*Proof:* See Appendix D. ∎

*Remark 4 - Global convergence and robustness of the algorithm.* The *global* convergence of the nonlinear asynchronous MIMO IWFA is guaranteed under condition (C1) that, differently from [16] where only square nonsingular channel matrices were considered, is valid for arbitrary channel matrices, either tall/fat or singular.

Observe that Algorithm 1 contains as special cases a plethora of algorithms, each one obtained by a possible choice of the scheduling of the users in the updating procedure (i.e., the parameters $\{\tau_r^q(n)\}$ and $\{\mathcal{T}_q\}$). All these algorithms can be implemented in a distributed way, where each user, to maximize his own rate, only needs to measure the covariance matrix of the overall interference-plus-noise and waterfill over this matrix. Two well-known special cases are the *sequential* and the *simultaneous* MIMO IWFA, where the users update their own strategies *sequentially* and *simultaneously,* respectively [16]. The important result stated in Theorem 12 is that all the algorithms resulting as special cases of the asynchronous MIMO IWFA are guaranteed to reach the unique NE of the game, under the same set of convergence conditions, since conditions in (C1) do not depend on the particular choice of $\{\mathcal{T}_q\}$ and $\{\tau_r^q(n)\}$.

### B. The MIMO Iterative Projection Algorithm

So far we have seen that, in the case of (strictly) tall (or rank deficient) channel matrices, the conditions needed for the multiuser waterfilling mapping to be a contraction (and thus guaranteeing the uniqueness of the NE of $\mathscr{G}$) are stronger than those required in the case of full row-rank channels. Furthermore, the former conditions depend also on the power budgets of the users, whereas the latter depend only on the channels. The difference in the two cases above is due to the fact that the reverse order law of generalized inverses of the products of matrices (Lemma 3), which plays a fundamental rule in the proof of the contraction property of the **WF** in the case of full row-rank matrices (Theorem 5), does not hold in the case of (strictly) tall (or singular) matrices [see (23)].

We would like to have a unified set of conditions guaranteeing both the uniqueness of the NE and the convergence of asynchronous distributed algorithms, possibly independent of the users' transmit powers and valid for *all* channel structures. In this section we focus on this issue and propose a new game, based on a variation of the original waterfilling best-response mapping, having the desired properties: the uniqueness of the NE as well as the convergence of distributed algorithms to the NE are guaranteed under the same unified set of conditions that are valid for *all* channel matrices (fat/square/tall and singular/nonsingular) and coincide with those given for the game $\mathscr{G}$ in the case of full row-rank channel matrices (the weakest one). This leads to a new game with, in general, different Nash equilibria with respect to the original game $\mathscr{G}$, but with almost the same performance in terms of sum-rate. Interestingly, the two games coincide if all the (direct) channel matrices are square nonsingular (when Lemma 3 holds).

To describe the new game, we focus w.l.o.g. only on the case in which the direct channel matrices are full column-rank matrices. The rank deficient case can be cast in the full column-rank case (cf. Appendix C). The best-response mapping of the new game is based on the following simple idea, inspired to the proof of Theorem 5: to obtain the desired conditions for the uniqueness of the NE and the convergence of distributed algorithms, we would like to use as best-response of each user $q$ the projection of the (minus) RHS in (24) onto the feasible set $\mathscr{Q}_q$, even in the case in which the reverse order law of the inverse in (24) does not hold. This choice leads to the MIMO waterfilling solution $\mathbf{WF}_q(\cdot)$ in (9) if Lemma 3 holds true; otherwise it provides a different mapping that corresponds to a modified channel or, equivalently, modified game. Stated in mathematical terms, the proposed best response of user $q$, given the strategies of the others $\mathbf{Q}_{-q} \in \mathscr{Q}_{-q}$, is

$$\widetilde{\mathbf{WF}}_q(\mathbf{Q}_{-q}) \triangleq \left[-\left(\mathbf{H}_{qq}^{\sharp}\mathbf{R}_{-q}(\mathbf{Q}_{-q})\mathbf{H}_{qq}^{\sharp H}\right)\right]_{\mathscr{Q}_q}. \quad (78)$$

Interestingly, the proposed best-response mapping can be still interpreted as a waterfilling solution, but over a proper modified channel, as shown next. Invoking Lemma 1 [see also (11)], (78) can be rewritten as

$$\widetilde{\mathbf{WF}}_q(\mathbf{Q}_{-q}) = \left[-\left(\left(\mathbf{H}_{qq}^{\sharp}\mathbf{R}_{-q}(\mathbf{Q}_{-q})\mathbf{H}_{qq}^{\sharp H}\right)^{-1}\right)^{-1}\right]_{\mathscr{Q}_q} \quad (79)$$

$$= \mathbf{WF}_q\left(\left(\mathbf{H}_{qq}^{\sharp}\mathbf{R}_{-q}(\mathbf{Q}_{-q})\mathbf{H}_{qq}^{\sharp H}\right)^{-1}\right) \quad (80)$$

where, with a slight abuse of notation, we used the same notation as in (9) to denote the MIMO waterfilling solution over the channel $\left(\mathbf{H}_{qq}^{\sharp}\mathbf{R}_{-q}(\mathbf{Q}_{-q})\mathbf{H}_{qq}^{\sharp H}\right)^{-1}$. We call this new game based on the mapping in (80) as $\widetilde{\mathscr{G}} \triangleq \left\{\Omega, \{\mathscr{Q}_q\}_{q \in \Omega}, \{\widetilde{R}_q\}_{q \in \Omega}\right\}$, where $\Omega = \{1, \cdots, Q\}$, $\mathscr{Q}_q$ is defined in (6), and $\widetilde{R}_q(\mathbf{Q})$ is the payoff of user $q$, defined as

$$\widetilde{R}_q(\mathbf{Q}_q, \mathbf{Q}_{-q}) \triangleq \log \det \left(\mathbf{I} + (\mathbf{H}_{qq}^{\sharp}\mathbf{R}_{-q}(\mathbf{Q}_{-q})\mathbf{H}_{qq}^{\sharp H})^{-1}\mathbf{Q}_q\right). \quad (81)$$

It follows then that, in game $\widetilde{\mathscr{G}}$, each player, given the strategy of the others, still performs a MIMO waterfilling solution, but over the modified equivalent channel $\left(\mathbf{H}_{qq}^{\sharp}\mathbf{R}_{-q}(\mathbf{Q}_{-q})\mathbf{H}_{qq}^{\sharp H}\right)^{-1}$, which is in general different from the one used in the original game $\mathscr{G}$, since the reverse order law of the inverse does not hold when the channel matrices $\mathbf{H}_{qq}$ are (strictly) tall, i.e., $\left(\mathbf{H}_{qq}^{\sharp}\mathbf{R}_{-q}(\mathbf{Q}_{-q})\mathbf{H}_{qq}^{\sharp H}\right)^{-1} \neq \mathbf{H}_{qq}^{H}\mathbf{R}_{-q}^{-1}(\mathbf{Q}_{-q})\mathbf{H}_{qq}$ [see (24)]. The two best-responses $\mathbf{WF}_q(\mathbf{Q}_{-q})$ in (9) and $\widetilde{\mathbf{WF}}_q(\mathbf{Q}_{-q})$ in (80) coincide if the channel matrices $\mathbf{H}_{qq}$ are square nonsingular.

The full characterization of game $\widetilde{\mathscr{G}}$ is given next.

*Theorem 13:* Game $\widetilde{\mathscr{G}}$ based on the mapping in (78) always admits a NE, for any set of channel matrices and transmit power of the users. Furthermore, suppose that the following condition holds true

$$\rho(\mathbf{S}) < 1, \quad \text{(C7)}$$

*where $\mathbf{S}$ is defined in* (20). *Then, the NE of the game is unique and, as $\mathrm{N}_{\mathrm{it}} \to \infty$, the asynchronous MIMO IWFA described in Algorithm 1 and based on the mapping in (78) converges to the unique NE of game $\widetilde{\mathscr{G}}$, for any set of feasible initial conditions and updating schedule.* □

## VI. NUMERICAL RESULTS

In this section, we provide some numerical results validating our theoretical findings. More specifically, we compare the performance of games $\mathscr{G}$ and $\widetilde{\mathscr{G}}$ in terms of conditions guaranteeing the uniqueness of the NE, achievable rates, and convergence speeds of some of the proposed algorithms.

*Example #1: Probability of the uniqueness of the NE and convergence of the asynchronous IWFA.* Since the conditions guaranteeing the uniqueness of the NE of games $\mathscr{G}$ and $\widetilde{\mathscr{G}}$ and convergence of the asynchronous IWFA given so far depend on the channel matrices $\{\mathbf{H}_{rq}\}_{r,q \in \Omega}$, there is a nonzero probability that they are not satisfied for a given channel realization drawn from a given probability space. To quantify the adequacy of our conditions, we tested them over a set of random channel matrices whose elements are generated as circularly symmetric complex Gaussian random variables with variance equal to the inverse of the square distance between the associated transmitter-receiver links (flat-fading channel model). We consider a MIMO multicell cellular network as depicted in Figure 1(a), composed of 7 (regular) hexagonal cells, sharing the same band. Hence, the transmissions from different cells typically interfere with each other. For the simplicity of representation, we assume that in each cell there is only one active MIMO link, corresponding to the transmission from the base station (BS) to a mobile terminal (MT). The overall network can be modeled as a 7-users interference channel, where each link is a MIMO channel. In Figure 1, we plot the probability that conditions (C1) (red line curves) and (C2) (blue line curves) are satisfied versus the (normalized) distance $d \in [0; 1)$ [see Figure 1(a)] between

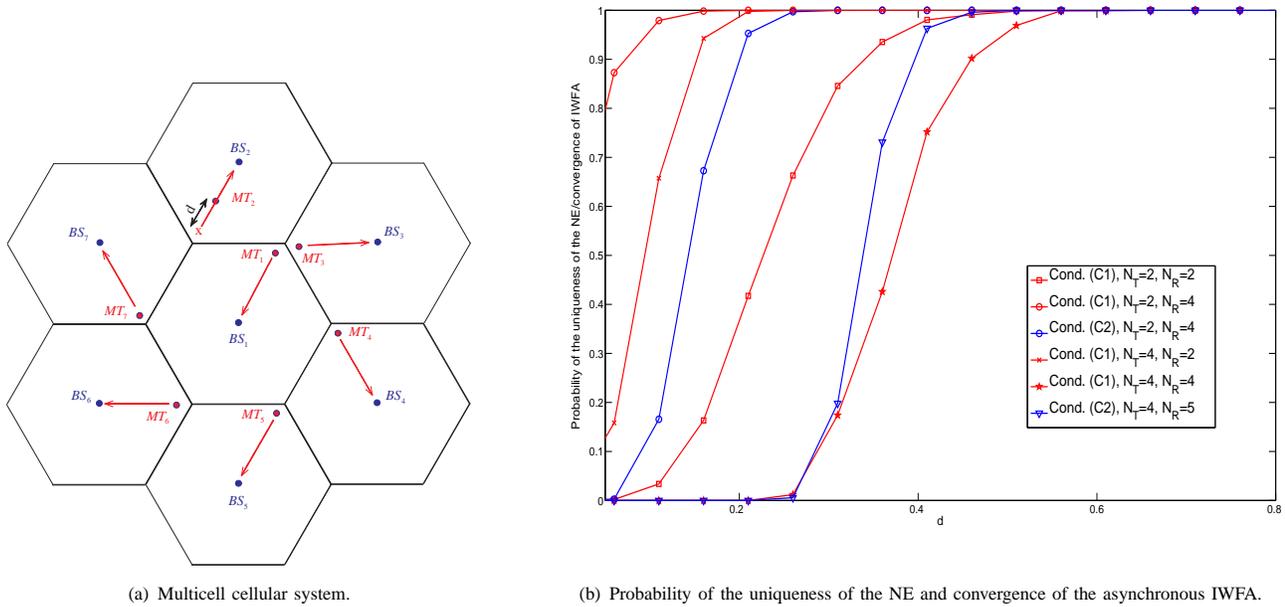

(a) Multicell cellular system.

(b) Probability of the uniqueness of the NE and convergence of the asynchronous IWFA.

Fig. 1. Probability of the uniqueness of the NE of games $\mathscr{G}$ and $\widetilde{\mathscr{G}}$ and convergence of the asynchronous IWFA; $Q = 7$, $n_{T_q} = n_{R_q} = 2$ (square marks), $n_{T_q} = 2$ and $n_{R_q} = 4$ (circle marks), $n_{T_q} = 4$ and $n_{R_q} = 2$ (cross marks), $n_{T_q} = n_{R_q} = 4$ (star marks), $n_{T_q} = 4$ and $n_{R_q} = 5$ (triangle marks), and $P_q/\sigma_q^2 = 5$dB, $\forall q \in \Omega$.

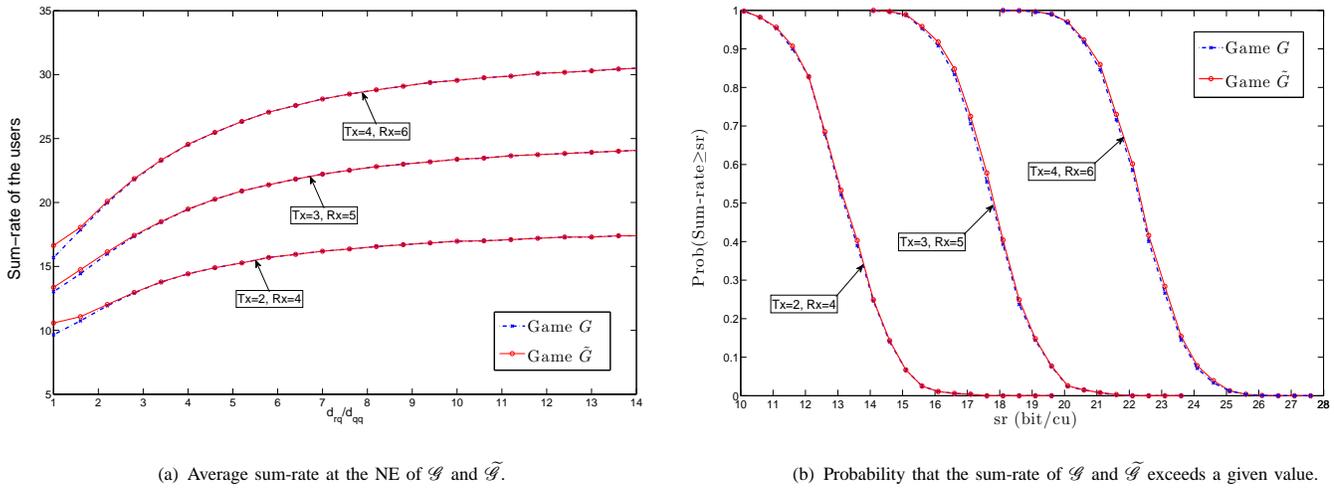

(a) Average sum-rate at the NE of $\mathscr{G}$ and $\widetilde{\mathscr{G}}$.

(b) Probability that the sum-rate of $\mathscr{G}$ and $\widetilde{\mathscr{G}}$ exceeds a given value.

Fig. 2. Performance of games $\mathscr{G}$ and $\widetilde{\mathscr{G}}$ in terms of Nash equilibria: Average sum-rate at the NE versus $d_{rq}/d_{qq}$ [subplot (a)] and probability that the sum-rate exceeds a given value [subplot (b)] versus the sum-rate values, for both games $\mathscr{G}$ (cross marks dashed-dot blue line curves) and $\widetilde{\mathscr{G}}$ (cross circle solid red line curves); $Q = 3$, $P_r = P_q$, $\text{snr}_q = P_q/\sigma_q^2 = 3$dB, $(n_{T_q}, n_{R_q}) \in \{(2,4),(3,5),(4,6)\}$ for all $r, q \in \Omega$.

each MT and his BS (assumed to be equal for all the MT/BS pairs), for different values of the transmit/receive antennas. We simulated (full rank) square, fat and tall channel matrices: $n_{T_q} = n_{R_q} = 2$ (square marks), $n_{T_q} = 2$ and $n_{R_q} = 4$ (circle marks), $n_{T_q} = 4$ and $n_{R_q} = 2$ (cross marks), $n_{T_q} = n_{R_q} = 4$ (star marks), $n_{T_q} = 4$ and $n_{R_q} = 5$ (triangle marks), for all $q = 1, \ldots, Q$. According to Theorems 9 and 13, we have the following interpretation for the probability curves plotted in Figure 1: if the channel matrices are (full rank) square or fat, then condition (C1) guarantees the uniqueness of the NE (and convergence of the IWFAs) of game $\mathscr{G}$; in the case of (full rank) tall channel matrices instead, condition (C1) applies to game $\widetilde{\mathscr{G}}$ whereas (C2) is valid for game $\mathscr{G}$. Looking at Figure 1 the following comments are in order:

- As expected, the probability of the uniqueness of the NE of both games $\widetilde{\mathscr{G}}$ and $\mathscr{G}$ and convergence of the IWFAs increases as each MT approaches his BS, corresponding to a decrease of the intercell interference;

- Sufficient condition (C2) is stronger than (C1) if computed over the *same* set of channels, as proved in (73) (see also Theorem 9 and Corollary 10). For example, in the case of full column-rank $2 \times 4$ MIMO channels, Figure 1 shows that, for any given distance $d \in [0,1)$, the probability that condition (C2) holds true (circle-marks blue curve) is lower than that of condition (C1) (circle-marks red curve), implying that game $\widetilde{\mathscr{G}}$ admits weaker uniqueness/convergence conditions than those of the original game $\mathscr{G}$;[11]

- Increasing the number of antennas at both the transmitter and receiver side leads to a decrease of the uniqueness/convergence probabilities (at least for the setup considered in the figure). For example, in the case of square MIMO channels, the curve associated to condition (C1) in the case of $2 \times 2$ MIMO channels (red square-marks curve) leads to better probability values than that obtained in the case of $4 \times 4$ MIMO channel (star-marks red curve), for any given distance $d$. The same result can be observed also in the case of tall channel matrices: condition (C2) for $2 \times 4$ channel matrices (circle-marks blue curve) has an higher probability to be satisfied than the same condition in the case of $4 \times 5$ MIMO channels (triangle-marks blue curve).

*Example #2: Game $\mathscr{G}$ vs. game $\widetilde{\mathscr{G}}$.* In Figure 2, we compare the performance of games $\mathscr{G}$ and $\widetilde{\mathscr{G}}$ in terms of sum-rate. Specifically, in Figure 2(a), we plot the sum-rate at the (unique) NE of the games $\mathscr{G}$ and $\widetilde{\mathscr{G}}$ for a three-user flat fading MIMO system as a function of the inter-pair distance among the links, for different number of transmit/receive antennas. In Figure 2(b), we plot the probability that the sum-rate exceeds a given value as a function of the sum-rate values, for the same systems as in Figure 2(a). The curves in both pictures are averaged over 1000 independent channel realizations, simulated as i.i.d. Gaussian random variables with zero mean and unit variance. For the sake of simplicity, the system is assumed to be symmetric, i.e., the transmitters have the same power budget and the interference links are at the same distance (i.e., $d_{rq} = d_{qr}$, $\forall q, r$), so that the cross channel gains are comparable in average sense. The path loss $\gamma$ is assumed to be $\gamma = 2.5$.

From the figures one infers that, as for isolated single-user systems or multiple access/broadcast channels, also in MIMO interference channels, increasing the number of antennas at both the transmitter and the receiver side leads to better performance. Furthermore, more interestingly, the figures show that the games $\mathscr{G}$ and $\widetilde{\mathscr{G}}$ have almost the same performance in terms of sum-rate at the NE, even if in the game $\widetilde{\mathscr{G}}$, given the strategies of the others, each player does not maximize his own rate, as instead he does in the game $\mathscr{G}$. This is due to the fact that the Nash equilibria of game $\mathscr{G}$ are in general not Pareto optimal, since pursuing the maximization of each individual objective function in a selfish manner does not guarantee in general the achievement of global optimality.

*Example #3: Simultaneous MIMO IWFA vs. sequential MIMO IWFA.* In Figure 3, we compare the performance of the sequential and simultaneous IWFAs applied to game $\mathscr{G}$, in terms of convergence speed, for a multicell cellular system composed of 7 (regular) hexagonal cells, as considered in the Example #1 [Figure 1(a)]. Here, we assume that in each cell there is only one active link, corresponding to the transmission from the base station (placed at the center of the cell) to a mobile terminal placed at (normalized) distance $d = 0.3$ from the corner of the cell [see Figure 1(a)]. In Figure 3, we show the rate evolution of the links of three cells corresponding to the sequential IWFA and simultaneous IWFA as a function of the iteration index and averaged on 500 independent channel realizations, each of them simulated as FIR filter of order $L = 6$. The curves refer to both cases of $n_{T_q} = n_{R_q} = 2$ and $n_{T_q} = n_{R_q} = 4$ for all links. To make the figure not excessively overcrowded, we report only the curves of 3 out of 7 links. As expected, the sequential IWFA is slower than the simultaneous IWFA, since each user is forced to wait for all the users scheduled in advance, before updating his own covariance matrix.

---

[11]Recall that, in the case of full column-rank channel matrices, condition (C1) is meaningless for game $\mathscr{G}$ and can be applied only to game $\widetilde{\mathscr{G}}$.

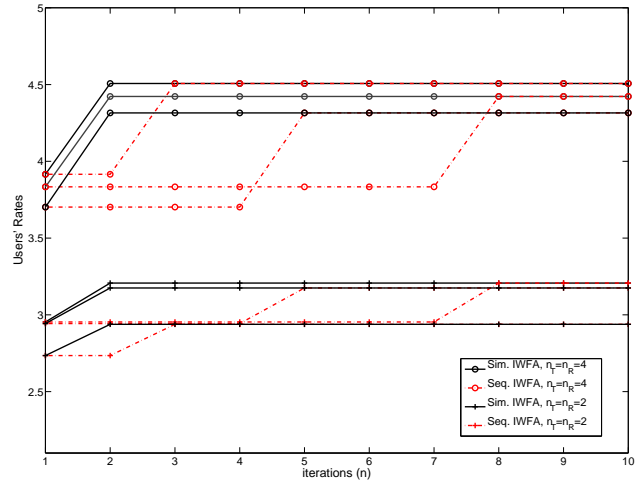

Fig. 3. Simultaneous IWFA (solid line curves) and sequential IWFA (dashed-dot line curves) vs. iterations [subplot (b)] for a 7 cell (downlink) wideband cellular system [subplot (a)]; Game $\mathscr{G}$, $n_{T_q} = n_{R_q} = 2$ (plus marks) and $n_{T_q} = n_{R_q} = 4$ (circle marks), $Q = 7$, 16 subcarriers, $P_q/\sigma_q^2 = 7$dB, $\forall r, q \in \Omega$.

## VII. CONCLUSIONS

In this paper we have considered a game theoretical formulation of the maximization of mutual information on each link, subject to power constraints, in the MIMO Gaussian interference channel. We have provided a complete characterization of the game, by deriving sufficient conditions guaranteeing both uniqueness of the NE and global convergence of the proposed totally asynchronous MIMO IWFA. Differently from current works in the literature, our results do not make any assumption on the structure of the channel matrices, but can be applied to *arbitrary* MIMO interference systems. Surprisingly, the uniqueness and convergence conditions in the case of (strictly) tall (and/or singular) channel matrices need to be stronger that those required in the case of full row-rank channel matrices. Finally, we have proposed an alternative algorithm, derived from a modified game, with weaker convergence conditions and virtually the same performance as the MIMO asynchronous IWFA.

# APPENDIX A
## PROOF OF LEMMA 1

We first introduce the following intermediate result and then proceed to the proof of Lemma 1.

*Lemma 14:* Let $\mathbf{X}_0 \in \mathbb{C}^{n \times n}$ be a Hermitian matrix with eigendecomposition $\mathbf{X}_0 = \mathbf{U}_0 \mathbf{D}_0 \mathbf{U}_0^H$, and let $\mathscr{Q}$ be the convex set defined as

$$\mathscr{Q} \triangleq \left\{ \mathbf{Q} \in \mathbb{C}^{n \times n} : \mathbf{Q} \succeq \mathbf{0}, \ \operatorname{Tr}\{\mathbf{Q}\} = P_T \right\}. \quad (82)$$

The matrix projection of $\mathbf{X}_0$ with respect to the Frobenius norm onto $\mathscr{Q}$, denoted by $[\mathbf{X}_0]_{\mathscr{Q}}$, is by definition the solution to the following convex optimization problem:

$$\begin{array}{ll} \underset{\mathbf{X}}{\text{minimize}} & \|\mathbf{X} - \mathbf{X}_0\|_F^2 \\ \text{subject to} & \mathbf{X} \in \mathscr{Q}, \end{array} \quad (83)$$

and takes the following form:

$$[\mathbf{X}_0]_{\mathscr{Q}} = \mathbf{U}_0 \left(\mathbf{D}_0 - \mu \mathbf{I}\right)^+ \mathbf{U}_0^H, \quad (84)$$

where $\mu$ satisfies the constraint $\operatorname{Tr}\{(\mathbf{D}_0 - \mu \mathbf{I})^+\} = P_T$.

*Proof:* Using $\mathbf{X}_0 = \mathbf{U}_0 \mathbf{D}_0 \mathbf{U}_0^H$, we rewrite the objective function of (83) as

$$\|\mathbf{X} - \mathbf{X}_0\|_F^2 = \left\|\widetilde{\mathbf{X}} - \mathbf{D}_0\right\|_F^2, \quad (85)$$

where we used the unitary invariance of the Frobenius norm [26] and $\widetilde{\mathbf{X}}$ is defined as $\widetilde{\mathbf{X}} \triangleq \mathbf{U}_0^H \mathbf{X} \mathbf{U}_0$. Since

$$\left\|\widetilde{\mathbf{X}} - \mathbf{D}_0\right\|_F^2 \geq \left\|\operatorname{Diag}\left(\widetilde{\mathbf{X}}\right) - \mathbf{D}_0\right\|_F^2, \quad (86)$$

with equality if and only if $\widetilde{\mathbf{X}}$ is diagonal, and the power constraint $\operatorname{Tr}\{\mathbf{X}\} = \operatorname{Tr}\{\widetilde{\mathbf{X}}\} = P_T$ depends only on the diagonal elements of $\widetilde{\mathbf{X}}$, it follows that the optimal $\widetilde{\mathbf{X}}$ must be diagonal: $\widetilde{\mathbf{X}} = \mathbf{D}$. Denoting by $d_k = [\mathbf{D}]_{kk}$ and $d_{k,0} = [\mathbf{D}_0]_{kk}$, with $k = 1, \cdots, n$, the diagonal entries of matrix $\mathbf{D}$ and $\mathbf{D}_0$, respectively, the matrix-valued problem (83) reduces to the following vector convex optimization problem

$$\begin{array}{ll} \underset{\mathbf{d} \geq \mathbf{0}}{\text{minimize}} & \sum_{k=1}^{n} (d_k - d_{0,k})^2 \\ \text{subject to} & \sum_{k=1}^{n} d_k = P_T, \end{array} \quad (87)$$

whose unique solution $\{d_k^\star\}$ is given by [15, Lemma 1]: $d_k^\star = (d_{0,k} - \mu)^+$, with $k = 1, \cdots, n$, where $\mu$ is chosen to satisfy $\sum_{k=1}^{n} (d_{0,k} - \mu)^+ = P_T$. ∎

*Corollary 15:* The Euclidean projection of $-\mathbf{X}_0$ onto the set $\mathscr{Q}$ defined in (82) takes the following form:

$$[-\mathbf{X}_0]_{\mathscr{Q}} = \mathbf{U}_0 \left(\mu \mathbf{I} - \mathbf{D}_0\right)^+ \mathbf{U}_0^H, \quad (88)$$

where $\mu$ is chosen to satisfy the constraint $\operatorname{Tr}\{(\mu \mathbf{I} - \mathbf{D}_0)^+\} = P_T$. □

*Proof of Lemma 1*: Given $q \in \Omega$ and $\mathbf{Q}_{-q} \in \mathscr{Q}_{-q}$, using the eigendecomposition of $\mathbf{H}_{qq}^H \mathbf{R}_{-q}^{-1}(\mathbf{Q}_{-q})\mathbf{H}_{qq}$ as given in (7), the Moore-Penrose pseudoinverse of $\mathbf{H}_{qq}^H \mathbf{R}_{-q}^{-1}(\mathbf{Q}_{-q})\mathbf{H}_{qq}$ can be written as [26] (we omit the dependence of $\mathbf{Q}_{-q}$ for the sake of notation):

$$\left(\mathbf{H}_{qq}^H \mathbf{R}_{-q}^{-1} \mathbf{H}_{qq}\right)^\sharp = \mathbf{U}_q \mathbf{D}_q^{-1} \mathbf{U}_q^H. \quad (89)$$

Using (89) and introducing the $n_{T_q} \times n_{T_q}$ unitary matrix $\tilde{\mathbf{U}}_q \triangleq \left(\mathbf{U}_q, \mathbf{U}_q^\perp\right)$, where $\mathbf{U}_q^\perp \in \mathbb{C}^{n_{T_q} \times n_{T_q} - r_q}$ is such that $\mathbf{U}_q^H \mathbf{U}_q^\perp = \mathbf{0}$, we have, for any given $c_q \in \mathbb{R}$:

$$\left(\mathbf{H}_{qq}^H \mathbf{R}_{-q}^{-1} \mathbf{H}_{qq}\right)^\sharp + c_q \mathbf{P}_{\mathcal{N}(\mathbf{H}_{qq})} = \tilde{\mathbf{U}}_q \begin{pmatrix} \mathbf{D}_q^{-1} & \mathbf{0} \\ \mathbf{0} & c_q \mathbf{I}_{n_{T_q} - r_q} \end{pmatrix} \tilde{\mathbf{U}}_q^H, \quad (90)$$

where $\mathbf{I}_{n_{T_q} - r_q}$ denotes the $(n_{T_q} - r_q) \times (n_{T_q} - r_q)$ identity matrix and the equality in (90) comes from $\mathbf{P}_{\mathcal{N}(\mathbf{H}_{qq})} = \mathbf{U}_q^\perp \mathbf{U}_q^{\perp H}$, implied from the fact that $\mathcal{N}(\mathbf{H}_{qq}^H \mathbf{R}_{-q}^{-1} \mathbf{H}_{qq}) = \mathcal{N}(\mathbf{H}_{qq})$ [recall that $\mathbf{R}_{-q}$ is positive definite for any $\mathbf{Q}_{-q} \in \mathscr{Q}_{-q}$]. It follows from Corollary 15 that, for any given $c_q \in \mathbb{R}_{++}$,

$$\left[-\left(\left(\mathbf{H}_{qq}^H \mathbf{R}_{-q}^{-1} \mathbf{H}_{qq}\right)^\sharp + c_q \mathbf{P}_{\mathcal{N}(\mathbf{H}_{qq})}\right)\right]_{\mathscr{Q}_q} = \tilde{\mathbf{U}}_q \left(\mu_q \mathbf{I} - \begin{pmatrix} \mathbf{D}_q^{-1} & \mathbf{0} \\ \mathbf{0} & c_q \mathbf{I}_{n_{T_q} - r_q} \end{pmatrix}\right)^+ \tilde{\mathbf{U}}_q^H \quad (91)$$

where $\mu_q$ is chosen to satisfy the constraint $\operatorname{Tr}\left(\left(\mu_q \mathbf{I} - \operatorname{Diag}(\mathbf{D}_q^{-1}, c_q \mathbf{I}_{n_{T_q} - r_q})\right)^+\right) = P_q$. Since $P_q < \infty$, it is always possible to find a sufficiently large positive constant $c_q < \infty$, such that $(\mu_q - c_q)^+ = 0$, and thus the RHS of (91) becomes

$$\left[-\left(\left(\mathbf{H}_{qq}^H \mathbf{R}_{-q}^{-1} \mathbf{H}_{qq}\right)^\sharp + c_q \mathbf{P}_{\mathcal{N}(\mathbf{H}_{qq})}\right)\right]_{\mathscr{Q}_q} = \mathbf{U}_q \left(\mu_q \mathbf{I} - \mathbf{D}_q^{-1}\right)^+ \mathbf{U}_q^H \quad (92)$$

which coincides with the desired solution in (9). In words: For any given finite power budget $P_q$, it is always possible to find a positive constant $c_q$ sufficiently large so that the waterfilling solution (9) does not allocate any power along the eigenvectors $\mathbf{U}_q^\perp$. Exploring the structure of the power allocation in (92), it is not difficult to show that this happens if $c_q \geq c_q(\mathbf{Q}_{-q}) = P_q + \max_{i \in \{1, \ldots, r_q\}} [\mathbf{D}_q]_{ii}^{-1}$ (sufficient condition). A (finite) upper bound of $c_q(\mathbf{Q}_{-q})$ independent on $\mathbf{Q}_{-q}$ is, e.g.,

$$c_q(\mathbf{Q}_{-q}) < P_q + \frac{\rho(\mathbf{R}_{n_q}) + Q P_{\max} \max_r \rho(\mathbf{H}_{rq}^H \mathbf{H}_{rq})}{\lambda_{\min}(\mathbf{H}_{qq}^H \mathbf{H}_{qq})} < \infty, \quad (93)$$

where $P_{\max} \triangleq \max_q P_q$. ∎

# APPENDIX B
## PROOF OF LEMMA 6 AND THEOREM 7

In this section we prove Lemma 6, and provide some intermediate results used in the proof of Theorem 7 given in Section III-B3.

We go through the complex differential of the (complex valued) functions using the approach in [32], meaning that we treat the complex differential of the complex variable and its complex conjugate as independent. In fact, it follows from the definition of complex derivative[12] that, although the complex variables $z \in \mathbb{C}$ (or $\mathbf{Z} \in \mathbb{C}^{m \times n}$) and $z^*$ (or $\mathbf{Z}^*$) are related, their derivatives (or differentials) are independent (or linear independent in the sense of [32, Lemma 1]). This approach simplifies the derivation of many complex derivative expressions.

*Definition 1:* Given $(\mathcal{D}_1, \mathcal{D}_2) \subseteq \mathbb{C}^{m \times n} \times \mathbb{C}^{m \times n}$, let $\mathbf{F}(\mathbf{X}_1, \mathbf{X}_2) : \mathcal{D}_1 \times \mathcal{D}_2 \mapsto \mathbb{C}^{p \times q}$ *be a complex matrix-valued function. Let* $(\mathbf{X}_1, \mathbf{X}_2)$ *be an interior point in* $\mathcal{D}_1 \times \mathcal{D}_2$, *and let* $(d\mathbf{X}_1, d\mathbf{X}_2)$ *such that* $(\mathbf{X}_1 + d\mathbf{X}_1, \mathbf{X}_2 + d\mathbf{X}_2)$ *lies in* $\mathcal{D}_1 \times \mathcal{D}_2$. *Then,* $\mathbf{F}(\mathbf{X}_1, \mathbf{X}_2)$ *is differentiable with respect to its first and second argument at* $(\mathbf{X}_1, \mathbf{X}_2)$ *if*

$$\begin{aligned}\mathbf{F}(\mathbf{X}_1 + d\mathbf{X}_1, \mathbf{X}_2 + d\mathbf{X}_2) &= \mathbf{F}(\mathbf{X}_1, \mathbf{X}_2) \\ &+ d\mathbf{F}(\mathbf{X}_1, \mathbf{X}_2; d\mathbf{X}_1, d\mathbf{X}_2) + o(\mathbf{X}_1, \mathbf{X}_2, d\mathbf{X}_1, d\mathbf{X}_2)\end{aligned} \quad (94)$$

*where* $d\mathbf{F}(\mathbf{X}_1, \mathbf{X}_2; d\mathbf{X}_1, d\mathbf{X}_2)$ *is a linear function in* $(d\mathbf{X}_1, d\mathbf{X}_2)$ *and* $o(\mathbf{X}_1, \mathbf{X}_2, d\mathbf{X}_1, d\mathbf{X}_2)$ *contains higher order terms of* $(d\mathbf{X}_1, d\mathbf{X}_2)$. $\square$

Definition 1 is the generalization of the analogous definition given in [31, Ch 5.4] and [33] for the real case to complex matrix-valued functions (see also [32]). Building on this definition, similarly to the real case, one can define the Jacobian matrices associated to a differentiable complex matrix-valued function. As in [32], we will consider functions $\mathbf{F}(\mathbf{X}, \mathbf{X}^*)$ that depend on both $\mathbf{X}$ and $\mathbf{X}^*$.

*Definition 2:* Given $\mathcal{D} \subseteq \mathbb{C}^{m \times n}$, *let* $\mathbf{F}(\mathbf{X}, \mathbf{X}^*) : \mathcal{D} \times \mathcal{D} \mapsto \mathbb{C}^{p \times q}$ *be a complex matrix-valued function, assumed to be differentiable on the interior of* $\mathcal{D} \times \mathcal{D}$. *The* $pq \times mn$ *Jacobian matrices of* $\mathbf{F}$ *with respect to* $\mathbf{X}$ *and* $\mathbf{X}^*$, *denoted by* $\mathbf{D_X F}(\mathbf{X}, \mathbf{X}^*)$ *and* $\mathbf{D_{X^*} F}(\mathbf{X}, \mathbf{X}^*)$, *respectively, are implicitly defined by the following differential expression [32]:*

$$\text{dvec}(\mathbf{F}) = (\mathbf{D_X F})\,\text{dvec}(\mathbf{X}) + (\mathbf{D_{X^*} F})\,\text{dvec}(\mathbf{X}^*). \quad (95)$$

$\square$

Using the notion of partial derivative, the Jacobian matrices $\mathbf{D_X F}(\mathbf{X}, \mathbf{X}^*)$ and $\mathbf{D_{X^*} F}(\mathbf{X}, \mathbf{X}^*)$ in Definition 2 can be written as

$$\begin{aligned}\mathbf{D_X F}(\mathbf{X}, \mathbf{X}^*) &= \frac{\partial \text{vec}(\mathbf{F}(\mathbf{X}, \mathbf{X}^*))}{\partial \text{vec}^T(\mathbf{X})}, \\ \mathbf{D_{X^*} F}(\mathbf{X}, \mathbf{X}^*) &= \frac{\partial \text{vec}(\mathbf{F}(\mathbf{X}, \mathbf{X}^*))}{\partial \text{vec}^T(\mathbf{X}^*)}.\end{aligned} \quad (96)$$

In [32, Lemma 1], the authors proved that, as for the real case [31, Ch.5, Th.3], the representation in the form (95) is unique, implying that (95) provides the identification rule for computing $\mathbf{D_X F}(\mathbf{X}, \mathbf{X}^*)$ and $\mathbf{D_{X^*} F}(\mathbf{X}, \mathbf{X}^*)$. In fact, given $\mathbf{F}(\mathbf{X}, \mathbf{X}^*)$, one can obtain the Jacobian matrices from the differential $d\mathbf{F}$, using the following three-step procedure: 1) compute the differential of $\mathbf{F}(\mathbf{X}, \mathbf{X}^*)$; 2) vectorize to obtain $\text{dvec}(\mathbf{F}) = \mathbf{A}(\mathbf{X}, \mathbf{X}^*)\,\text{dvec}(\mathbf{X}) + \mathbf{B}(\mathbf{X}, \mathbf{X}^*)\,\text{dvec}(\mathbf{X}^*)$; and 3) conclude that $\mathbf{D_X F}(\mathbf{X}, \mathbf{X}^*) = \mathbf{A}(\mathbf{X}, \mathbf{X}^*)$ and $\mathbf{D_{X^*} F}(\mathbf{X}, \mathbf{X}^*) = \mathbf{B}(\mathbf{X}, \mathbf{X}^*)$.

Using the above definitions, we can now prove the following.

*Lemma 16:* Let $\mathbf{F}(\mathbf{X}) : \mathcal{D} \subseteq \mathbb{C}^{m \times n} \mapsto \mathbb{C}^{p \times q}$ *be a complex matrix-valued function, assumed to be differentiable at an interior point* $\mathbf{X}_0$ *in the set* $\mathcal{D}$, *with Jacobian matrix at* $\mathbf{X}_0$ *denoted by* $\mathbf{D_X F}(\mathbf{X}_0)$. *Given* $\mathbf{x} \triangleq [\text{vec}^T(\text{Re}\{\mathbf{X}\}), \text{vec}^T(\text{Im}\{\mathbf{X}\})]^T$, *let* $\mathbf{f}(\mathbf{x}) : \mathbb{R}^{2mn} \mapsto \mathbb{R}^{2pq}$ *be the real vector-valued function defined as* $\mathbf{f}(\mathbf{x}) \triangleq [\text{vec}^T(\text{Re}\,\mathbf{F}(\{\mathbf{X}\})), \text{vec}^T(\text{Im}\,\mathbf{F}(\{\mathbf{X}\}))]^T$. *Then,* $\mathbf{f}$ *is differentiable at* $\mathbf{x}_0$, *with Jacobian matrix given by*

$$\begin{aligned}\mathbf{D_x f}(\mathbf{x}_0) &\triangleq \left.\frac{\partial \text{vec}(\mathbf{f}(\mathbf{x}))}{\partial \text{vec}^T(\mathbf{x})}\right|_{\mathbf{x}=\mathbf{x}_0} \\ &= \begin{bmatrix} \text{Re}\{\mathbf{D_X F}(\mathbf{X}_0)\} & -\text{Im}\{\mathbf{D_X F}(\mathbf{X}_0)\} \\ \text{Im}\{\mathbf{D_X F}(\mathbf{X}_0)\} & \text{Re}\{\mathbf{D_X F}(\mathbf{X}_0)\}\end{bmatrix}\end{aligned} \quad (97)$$

*Proof:* The differentiability of $\mathbf{f}$ follows directly from that of $\mathbf{F}$. Given $\mathbf{X} = \mathbf{X}_R + j\mathbf{X}_I$, with $\mathbf{X}_R, \mathbf{X}_I \in \mathbb{R}^{m \times n}$, the Jacobian matrix $\mathbf{D_x f}(\mathbf{x}_0)$ of $\mathbf{f}$ at $\mathbf{x}_0$ is [see (96)]:

$$\mathbf{D_x f}(\mathbf{x}_0) = \begin{bmatrix} \mathbf{D}_{\mathbf{X}_R} \text{Re}\{\mathbf{F}(\mathbf{X}_0)\} & \mathbf{D}_{\mathbf{X}_I} \text{Re}\{\mathbf{F}(\mathbf{X}_0)\} \\ \mathbf{D}_{\mathbf{X}_R} \text{Im}\{\mathbf{F}(\mathbf{X}_0)\} & \mathbf{D}_{\mathbf{X}_I} \text{Im}\{\mathbf{F}(\mathbf{X}_0)\}\end{bmatrix}. \quad (98)$$

We compute now the four matrix blocks in $\mathbf{D_x f}(\mathbf{x}_0)$ above. Invoking the chain rule for Jacobian matrices [32, Theorem 1] and using the following simple rules $\mathbf{D}_{\mathbf{X}_R} \mathbf{X} = \mathbf{D}_{\mathbf{X}_R} \mathbf{X}^* = \mathbf{I}_{mn}, \mathbf{D}_{\mathbf{X}_I} \mathbf{X} = -\mathbf{D}_{\mathbf{X}_I} \mathbf{X}^* = j\mathbf{I}_{mn}$, we obtain:

$$\begin{aligned}\mathbf{D}_{\mathbf{X}_R} \text{Re}\{\mathbf{F}(\mathbf{X})\} &= 2\,\text{Re}\{\mathbf{D_X}\,\text{Re}\{\mathbf{F}(\mathbf{X})\}\}, \quad &(99)\\ \mathbf{D}_{\mathbf{X}_I} \text{Re}\{\mathbf{F}(\mathbf{X})\} &= -2\,\text{Im}\{\mathbf{D_X}\,\text{Re}\{\mathbf{F}(\mathbf{X})\}\}, \quad &(100)\\ \mathbf{D}_{\mathbf{X}_R} \text{Im}\{\mathbf{F}(\mathbf{X})\} &= 2\,\text{Re}\{\mathbf{D_X}\,\text{Im}\{\mathbf{F}(\mathbf{X})\}\}, \quad &(101)\\ \mathbf{D}_{\mathbf{X}_I} \text{Im}\{\mathbf{F}(\mathbf{X})\} &= -2\,\text{Im}\{\mathbf{D_X}\,\text{Im}\{\mathbf{F}(\mathbf{X})\}\}. \quad &(102)\end{aligned}$$

Given (99)-(102), to complete the proof, we need to compute $\mathbf{D_X}\,\text{Re}\{\mathbf{F}(\mathbf{X})\}$ and $\mathbf{D_X}\,\text{Im}\{\mathbf{F}(\mathbf{X})\}$. Invoking the identification rule (95) in Definition 2, we have

$$\begin{aligned}\text{dvec}(\text{Re}\{\mathbf{F}\}) &= \frac{1}{2}\text{dvec}(\mathbf{F} + \mathbf{F}^*) \\ &= \left(\frac{1}{2}\mathbf{D_X F}\right)\text{dvec}(\mathbf{X}) + \left(\frac{1}{2}\mathbf{D_X F}\right)^* \text{dvec}(\mathbf{X}^*) \\ &\triangleq (\mathbf{D_X}\,\text{Re}\{\mathbf{F}(\mathbf{X})\})\,\text{dvec}(\mathbf{X}) + (\mathbf{D_{X^*}}\,\text{Re}\{\mathbf{F}(\mathbf{X})\})\,\text{dvec}(\mathbf{X}^*),\end{aligned} \quad (103)$$

where we used the fact that $\mathbf{F} = \mathbf{F}(\mathbf{X})$ depends only on $\mathbf{X}$ (and not on $\mathbf{X}^*$), and

$$\begin{aligned}\mathbf{D_X}\,\text{Re}\{\mathbf{F}(\mathbf{X})\} &= \frac{1}{2}\mathbf{D_X F}, \quad &(104)\\ \mathbf{D_{X^*}}\,\text{Re}\{\mathbf{F}(\mathbf{X})\} &= \frac{1}{2}(\mathbf{D_X F})^*. \quad &(105)\end{aligned}$$

Similarly, we can compute

$$\begin{aligned}\mathbf{D_X}\,\text{Im}\{\mathbf{F}(\mathbf{X})\} &= \frac{1}{2j}\mathbf{D_X F}, \quad &(106)\\ \mathbf{D_{X^*}}\,\text{Im}\{\mathbf{F}(\mathbf{X})\} &= -\frac{1}{2j}(\mathbf{D_X F})^*. \quad &(107)\end{aligned}$$

---

[12] Given $f : \mathbb{C} \mapsto \mathbb{C}$ and $x = x_R + jx_I$, with $x_R, x_I \in \mathbb{R}$, the partial derivatives (or Wirtinger derivatives [33]) $\frac{\partial}{\partial x}f(x_0)$ and $\frac{\partial}{\partial x^*}f(x_0)$ are defined as follows: $\frac{\partial}{\partial x}f(x_0) \triangleq \frac{1}{2}\left(\frac{\partial}{\partial x_R}f(x) - j\frac{\partial}{\partial x_I}f(x)\right)\bigg|_{x=x_0}$ and $\frac{\partial}{\partial x^*}f(x_0) \triangleq \frac{1}{2}\left(\frac{\partial}{\partial x_R}f(x) + j\frac{\partial}{\partial x_I}f(x)\right)\bigg|_{x=x_0}$. Observe that in $\frac{\partial}{\partial x}f(x)$ and $\frac{\partial}{\partial x^*}f(x)$ the variables $x$ and $x^*$, respectively, are treated as independent variables.

$$\|\mathbf{D_x f}(\mathbf{z}_t)(\mathbf{y}-\mathbf{x})\|_2 = \left\| \begin{bmatrix} \operatorname{Re}\{\mathbf{D_X F}(\mathbf{Z}_t)\} & -\operatorname{Im}\{\mathbf{D_X F}(\mathbf{Z}_t)\} \\ \operatorname{Im}\{\mathbf{D_X F}(\mathbf{Z}_t)\} & \operatorname{Re}\{\mathbf{D_X F}(\mathbf{Z}_t)\} \end{bmatrix} \begin{bmatrix} \operatorname{vec}(\operatorname{Re}\{\mathbf{Y}-\mathbf{X}\}) \\ \operatorname{vec}(\operatorname{Im}\{\mathbf{Y}-\mathbf{X}\}) \end{bmatrix} \right\|_2 \quad (115)$$

$$= \|[\operatorname{Re}\{\mathbf{D_X F}(\mathbf{Z}_t)\} + j\operatorname{Im}\{\mathbf{D_X F}(\mathbf{Z}_t)\}][\operatorname{vec}(\operatorname{Re}\{\mathbf{Y}-\mathbf{X}\}) + j\operatorname{vec}(\operatorname{Im}\{\mathbf{Y}-\mathbf{X}\})]\|_2 \quad (116)$$

$$= \|\mathbf{D_X F}(\mathbf{Z}_t)\operatorname{vec}(\mathbf{Y}-\mathbf{X})\|_2 \quad (117)$$

$$\leq \|\mathbf{D_X F}(\mathbf{Z}_t)\|_{2,\operatorname{mat}}\|\mathbf{Y}-\mathbf{X}\|_F, \quad (118)$$

---

Introducing (104)-(107) in (99)-(102) and using (98) we obtain the desired expression for $\mathbf{D_x f(x)}$ as given in (97). This completes the proof. ∎

We have now all the result we need to prove the mean-value theorem for complex matrix-valued functions.

*Proof of Lemma 6*: We organize the proof in two steps. First, we apply the classical mean-value theorem given in (44) [34, Th.5.10] to a proper *real scalar* function $\phi(t)$, related to the original complex matrix-valued function $\mathbf{F(X)}$. Then, using Lemma 16, we convert the obtained result in the form (46)-(47).

*Step 1:* Given $\mathbf{X}$ and $\mathbf{F(X)}$, let $\mathbf{x} \triangleq [\operatorname{vec}^T(\operatorname{Re}\{\mathbf{X}\}), \operatorname{vec}^T(\operatorname{Im}\{\mathbf{X}\})]^T$ and $\mathbf{f(x)}: \mathbb{R}^{2mn} \mapsto \mathbb{R}^{2pq}$ be the real vector-valued function defined as $\mathbf{f(x)} \triangleq [\operatorname{vec}^T(\operatorname{Re}\mathbf{F}(\{\mathbf{X}\})), \operatorname{vec}^T(\operatorname{Im}\mathbf{F}(\{\mathbf{X}\}))]^T$. It follows directly from the assumptions on $\mathbf{F}$, that $\mathbf{f}$ is continuous (and differentiable) in (the interior of) its domain. Given $\mathbf{X}, \mathbf{Y} \in \mathcal{D}$, with $\mathbf{X} \neq \mathbf{Y}$, we introduce the real scalar function $\phi: [0,1] \mapsto \mathbb{R}$, defined as

$$\phi(t) \triangleq \frac{<\mathbf{f(y)}-\mathbf{f(x)}, \mathbf{f}(t\mathbf{y}+(1-t)\mathbf{x})>}{\|\mathbf{f(y)}-\mathbf{f(x)}\|_2}, \quad t\in[0,1], \quad (108)$$

where $<\mathbf{x},\mathbf{y}> \triangleq \mathbf{x}^T\mathbf{y}$ and $\|\cdot\|_2$ is the Euclidean norm. Observe that the function $\phi(t)$ is well-defined because the set $\mathcal{D}$ is convex. Furthermore, $\phi(t)$ is continuous on $[0,1]$ and differentiable on $(0,1)$, implied from the differentiability of $\mathbf{f}$ and the scalar product. The first derivative of $\phi(t)$ is equal to:

$$\frac{d\phi(t)}{dt} = \frac{<\mathbf{f(y)}-\mathbf{f(x)}, \mathbf{D_x f}(t\mathbf{y}+(1-t)\mathbf{x})(\mathbf{y}-\mathbf{x})>}{\|\mathbf{f(y)}-\mathbf{f(x)}\|_2}, \quad (109)$$

where in (**??**) we used the chain rule (see, e.g., [31, Ch. 5, Theorem 8]).

Function $\phi(t)$ satisfies the conditions of the mean value theorem for real scalar functions [34, Th.5.10], meaning that [see (44)]

$$\phi(1)-\phi(0) = \left.\frac{d\phi(z)}{dz}\right|_{z=t}, \quad \text{for some } t\in(0,1). \quad (110)$$

Using (**??**) and

$$\phi(1) \triangleq \frac{<\mathbf{f(y)}-\mathbf{f(x)}, \mathbf{f(y)}>}{\|\mathbf{f(y)}-\mathbf{f(x)}\|_2}, \quad (111)$$

$$\phi(0) \triangleq \frac{<\mathbf{f(y)}-\mathbf{f(x)}, \mathbf{f(x)}>}{\|\mathbf{f(y)}-\mathbf{f(x)}\|_2}, \quad (112)$$

the LHS in (110) becomes: $\phi(1)-\phi(0) = \|\mathbf{f(y)}-\mathbf{f(x)}\|_2$, whereas the RHS becomes

$$\left.\frac{d\phi(z)}{dz}\right|_{z=t} = \frac{<\mathbf{f(y)}-\mathbf{f(x)}, \mathbf{D_x f}(t\mathbf{y}+(1-t)\mathbf{x})(\mathbf{y}-\mathbf{x})>}{\|\mathbf{f(y)}-\mathbf{f(x)}\|_2}$$

$$\leq \|\mathbf{D_x f}(t\mathbf{y}+(1-t)\mathbf{x})(\mathbf{y}-\mathbf{x})\|_2, \quad (113)$$

which leads to

$$\|\mathbf{f(y)}-\mathbf{f(x)}\|_2 \leq \|\mathbf{D_x f}(t\mathbf{y}+(1-t)\mathbf{x})(\mathbf{y}-\mathbf{x})\|_2, \quad (114)$$

for some $t \in (0,1)$.

*Step 2:* To complete the proof, we need to show that (46) comes from (114). To this end, we use Lemma 16, as detailed next. Since $\|\mathbf{f(y)}-\mathbf{f(x)}\|_2 = \|\mathbf{F(Y)}-\mathbf{F(X)}\|_F$, we focus only on the RHS of (114). For the sake of notation, we introduce $\mathbf{Z}_t \triangleq t\mathbf{Y}+(1-t)\mathbf{X}$ and $\mathbf{z}_t \triangleq [\operatorname{vec}^T(\operatorname{Re}\{\mathbf{Z}_t\}), \operatorname{vec}^T(\operatorname{Im}\{\mathbf{Z}_t\})]^T$, so that the RHS of (114) can be written as in the equation (115) at the bottom of the page, where we used Lemma 16. The desired expressions in (46)-(47) follows directly from (114) and (117)-(118). This completes the proof. ∎

We provide also the following lemma that is used in the proof of Theorem 7.

*Lemma 17:* Let $\mathbf{F(X)} \triangleq (\mathbf{H}^H\mathbf{R}^{-1}(\mathbf{X})\mathbf{H})^{-1}$, with $\mathbf{R(X)} \triangleq \mathbf{C} + \sum_{k=1}^{K}\mathbf{T}_k\mathbf{X}_k\mathbf{T}_k^H$, where $\mathbf{H} \in \mathbb{C}^{m\times n}$, $\mathbf{C} \in \mathbb{C}^{m\times m}$, $\mathbf{T}_k \in \mathbb{C}^{m\times r_k}$, and $\mathbf{X} \triangleq [\mathbf{X}_1,\cdots\mathbf{X}_k]$ with each $\mathbf{X}_k \in \mathbb{C}^{r_k\times r_k}$, such that $\operatorname{rank}(\mathbf{H})=n$, $\operatorname{rank}(\mathbf{C})=m$, and $\mathbf{R(X)}$ is non singular. Then $\mathbf{F(X)}$ is differentiable at $\mathbf{X}$ and

$$\mathbf{D_X F(X)} = [\mathbf{G}_1^*(\mathbf{X}) \otimes \mathbf{G}_1(\mathbf{X}), \cdots, \mathbf{G}_K^*(\mathbf{X}) \otimes \mathbf{G}_K(\mathbf{X})] \quad (119)$$

where

$$\mathbf{G}_k(\mathbf{X}) = (\mathbf{H}^H\mathbf{R}^{-1}(\mathbf{X})\mathbf{H})^{-1}\mathbf{H}^H\mathbf{R}^{-1}(\mathbf{X})\mathbf{T}_k. \quad (120)$$

*Proof:* We compute first the differential of $\mathbf{F(X)}$ and then invoking the identification rule in (95) we identify the Jacobian matrix. Function $\mathbf{F}$ is differentiable at $\mathbf{X}$, with differential given by

$$d\operatorname{vec}\mathbf{F(X)}$$
$$= -(\mathbf{H}^H\mathbf{R}^{-1}(\mathbf{X})\mathbf{H})^{-1}\mathbf{H}^H d\mathbf{R}^{-1}(\mathbf{X})\mathbf{H}(\mathbf{H}^H\mathbf{R}^{-1}(\mathbf{X})\mathbf{H})^{-1}$$
$$\triangleq \sum_{k=1}^{K}\mathbf{G}_k(\mathbf{X})d\mathbf{X}_k\mathbf{G}_k^H(\mathbf{X}), \quad (121)$$

where we used $d\mathbf{X}^{-1} = -\mathbf{X}^{-1}d\mathbf{X}^{-1}\mathbf{X}^{-1}$ [32] and (120). By vectorizing $d\mathbf{F(X)}$ and using $\operatorname{vec}(\mathbf{ABC}) = (\mathbf{C}^T\otimes\mathbf{A})\operatorname{vec}(\mathbf{B})$ (see, e.g., [31], [26]) we obtain

$$d\operatorname{vec}\mathbf{F(X)}$$
$$= [\mathbf{G}_1^*(\mathbf{X})\otimes\mathbf{G}_1(\mathbf{X}), \ldots, \mathbf{G}_K^*(\mathbf{X})\otimes\mathbf{G}_K(\mathbf{X})]d\operatorname{vec}\mathbf{X}, \quad (122)$$

which, using the identification rule in (95) leads to the desired structure of the Jacobian $\mathbf{D_X F(X)}$ as given in (119). ∎

## APPENDIX C
## PROOF OF THEOREM 9

The existence of a NE of game $\mathscr{G}$ for any set of channel matrices and power budget follows directly from [16, Th. 6] (i.e., quasiconcave payoff functions and convex compact strategy sets). As far as the uniqueness of the NE is concerned, a sufficient condition for the uniqueness of the equilibrium is that the waterfilling mapping in (9) be a contraction with respect to some norm [23, Prop. 1.1.(a)]. Hence, the sufficiency of (C1) in the case of full row-rank and full column-rank channel matrices $\{\mathbf{H}_{qq}\}_{q\in\Omega}$ follows from Theorems 5 and 7, respectively. Finally, the equivalences $\|\mathbf{S}\|_{\infty,\mathrm{mat}}^{\mathbf{w}} < 1 \Leftrightarrow \rho(\mathbf{S}) < 1$ and $\|\mathbf{S}(\mathscr{P}^\star)\|_{\infty,\mathrm{mat}}^{\mathbf{w}} < 1 \Leftrightarrow \rho(\mathbf{S}(\mathscr{P}^\star)) < 1$ can be proved using [23, Cor. 6.1] (cf. [16, Th. 6]).

We focus now on the more general case in which the channel matrices $\mathbf{H}_{qq}$ may be rank deficient and prove that condition (C1) in Theorem 9 is still sufficient to guarantee the uniqueness of the NE. Let $\overline{\Omega} \subseteq \Omega$ be the subset of $\Omega$ containing the users' indexes $q$ such that $\mathrm{rank}(\mathbf{H}_{qq}) = r_q < \min(n_{T_q}, n_{R_q})$. For each $q \in \overline{\Omega}$, given the SVD of $\mathbf{H}_{qq} = \mathbf{U}_{q,1}\mathbf{\Sigma}_{qq}\mathbf{V}_{q,1}^H$, where $\mathbf{U}_{q,1} \in \mathbb{C}^{n_{R_q} \times r_q}$, $\mathbf{V}_{q,1} \in \mathbb{C}^{n_{T_q} \times r_q}$ are semi-unitary matrices and $\mathbf{\Sigma}_{qq} \in \mathbb{C}^{r_q \times r_q}$ is a diagonal matrices with positive entries, let $\mathbf{P}_{\mathcal{N}(\mathbf{H}_{qq})^\perp} = \mathbf{V}_{q,1}\mathbf{V}_{q,1}^H$ be the orthogonal projection onto the subspace orthogonal to the null-space of $\mathbf{H}_{qq}$. Since $r_q < \min(n_{T_q}, n_{R_q})$, it is not difficult to show that the best-response of each user $\mathbf{Q}_q^\star = \mathbf{WF}_q(\mathbf{Q}_{-q})$–the solution of the rate-maximization problem in (5) for a given $\mathbf{Q}_{-q} \in \mathscr{Q}_{-q}$–will be orthogonal to the null space of $\mathbf{H}_{qq}$, whatever $\mathbf{Q}_{-q} \in \mathscr{Q}_{-q}$ is, implying $\mathbf{Q}_q^\star = \mathbf{P}_{\mathcal{N}(\mathbf{H}_{qq})^\perp}\mathbf{Q}_q^\star\mathbf{P}_{\mathcal{N}(\mathbf{H}_{qq})^\perp}$. It follows then that the best response of each user $q \in \overline{\Omega}$ belongs to the following class of matrices:

$$\mathbf{Q}_q = \mathbf{V}_{q,1}\overline{\overline{\mathbf{Q}}}_q\mathbf{V}_{q,1}^H, \quad (123)$$

with

$$\overline{\overline{\mathbf{Q}}}_q \in \overline{\overline{\mathscr{Q}}}_q \triangleq \left\{\mathbf{X} \in \mathbb{C}^{r_q \times r_q} : \mathbf{X} \succeq \mathbf{0}, \quad \mathrm{Tr}(\mathbf{X}) = P_q\right\}. \quad (124)$$

Using (123) and introducing the (possibly) lower-dimensional covariance matrices $\{\overline{\mathbf{Q}}_q\}_{q\in\Omega}$ and the modified channel matrices $\{\overline{\mathbf{H}}_{rq}\}_{r,q\in\Omega}$, defined respectively as

$$\overline{\mathbf{Q}}_q \triangleq \begin{cases} \overline{\overline{\mathbf{Q}}}_q, & \text{if } q \in \overline{\Omega}, \\ \mathbf{Q}_q, & \text{otherwise,} \end{cases}$$
$$\overline{\mathbf{H}}_{rq} \triangleq \begin{cases} \mathbf{H}_{rq}\mathbf{V}_{r,1}, & \text{if } r \in \overline{\Omega}, \\ \mathbf{H}_{rq}, & \text{otherwise,} \end{cases} \quad (125)$$

game $\mathscr{G}$ can be recast in the following lower-dimensional game $\overline{\mathscr{G}}$, defined as

$$(\overline{\mathscr{G}}): \begin{array}{l}\underset{\overline{\mathbf{Q}}_q}{\text{maximize}} \quad \log\det\left(\mathbf{I} + \overline{\mathbf{H}}_{qq}^H \overline{\mathbf{R}}_{-q}^{-1} \overline{\mathbf{H}}_{qq}\overline{\mathbf{Q}}_q\right) \\ \text{subject to} \quad \overline{\mathbf{Q}}_q \in \overline{\mathscr{Q}}_q, \end{array} \quad \forall q \in \Omega, \quad (126)$$

where $\overline{\mathbf{R}}_{-q} = \overline{\mathbf{R}}_{-q}(\overline{\mathbf{Q}}_{-q}) \triangleq \mathbf{R}_{n_{R_q}} + \sum_{r \neq q}\overline{\mathbf{H}}_{rq}\overline{\mathbf{Q}}_r\overline{\mathbf{H}}_{rq}^H$; and $\overline{\mathscr{Q}}_q$ coincides with $\overline{\overline{\mathscr{Q}}}_q$ defined in (124) if $q \in \overline{\Omega}$, whereas it coincides with $\mathscr{Q}_q$ defined in (6) if $q \in \Omega\setminus\overline{\Omega}$. It is straightforward to see that conditions for the uniqueness of the NE of game $\overline{\mathscr{G}}$ guarantee also the uniqueness of the NE of $\mathscr{G}$.

Observe that, in the game $\overline{\mathscr{G}}$, all channel matrices $\overline{\mathbf{H}}_{qq}$ are full column-rank matrices. We can thus use Theorem 7 and obtain the following sufficient condition for the uniqueness of the NE of both games $\overline{\mathscr{G}}$ and $\mathscr{G}$:

$$\rho(\overline{\mathbf{S}}(\mathscr{P}^\star)) < 1, \quad (127)$$

with

$$\left[\overline{\mathbf{S}}(\mathscr{P}^\star)\right]_{qr}$$
$$\triangleq \begin{cases} \rho\left(\mathbf{V}_r^H\mathbf{H}_{rq}^H\mathbf{P}_{rq}^{\star H}\mathbf{H}_{qq}^{\sharp H}\mathbf{H}_{qq}^{\sharp}\mathbf{P}_{rq}^{\star}\mathbf{H}_{rq}\mathbf{V}_r\right), & \text{if } r \in \overline{\Omega},\, r \neq q, \\ \rho\left(\mathbf{H}_{rq}^H\mathbf{P}_{rq}^{\star H}\mathbf{H}_{qq}^{\sharp H}\mathbf{H}_{qq}^{\sharp}\mathbf{P}_{rq}^{\star}\mathbf{H}_{rq}\right), & \text{if } r \in \Omega\setminus\overline{\Omega},\, r \neq q, \\ 0 & \text{if } r = q, \end{cases} \quad (128)$$

with $\mathscr{P}^\star$ defined in Theorem 7 [recall that $\mathbf{P}_{rq}^\star = \mathbf{I}$ if $\mathrm{rank}(\mathbf{H}_{qq}) = n_{R_q}$, which includes the square nonsingular case], where in (128) we used $\rho\left(\overline{\mathbf{H}}_{rq}^H\mathbf{P}_{rq}^H\overline{\mathbf{H}}_{qq}^{\sharp H}\overline{\mathbf{H}}_{qq}^\sharp\mathbf{P}_{rq}\overline{\mathbf{H}}_{rq}\right) = \rho\left(\mathbf{V}_r^H\mathbf{H}_{rq}^H\mathbf{P}_{rq}^H\mathbf{H}_{qq}^{\sharp H}\mathbf{H}_{qq}^\sharp\mathbf{P}_{rq}\mathbf{H}_{rq}\mathbf{V}_r\right)$.

We show now that condition (C1) in Theorem 9 is sufficient for (127). Introducing the Hermitian positive semidefinite matrix $\mathbf{G}_{rq} \triangleq \mathbf{H}_{rq}^H\mathbf{P}_{rq}^H\mathbf{H}_{qq}^{\sharp H}\mathbf{H}_{qq}^\sharp\mathbf{P}_{rq}\mathbf{H}_{rq}$ and invoking the Poincar separation theorem [26, Cor. 4.3.16], we have for any given $\mathbf{P}_{rq} \in \mathscr{P}$:

$$\lambda_i(\mathbf{G}_{rq}) \leq \lambda_i(\mathbf{V}_r^H\mathbf{G}_{rq}\mathbf{V}_r) \leq \lambda_{i+n_{T_r}-r_r}(\mathbf{G}_{rq}), \quad (129)$$

$\forall i = 1, 2, \ldots, r_r$, where the eigenvalues $\{\lambda_i(\cdot)\}$ are arranged in increasing order. It follows from (129) that $\rho\left(\mathbf{V}_r^H\mathbf{G}_{rq}\mathbf{V}_r\right) \leq \rho(\mathbf{G}_{rq})$, for all $r \in \overline{\Omega}$, $q \in \Omega$ with $r \neq q$ and fixed $\mathbf{P}_{rq} \in \mathscr{P}$, and thus $\overline{\mathbf{S}}(\mathscr{P}^\star) \leq \mathbf{S}(\mathscr{P}^\star)$, where $\mathbf{S}$ is defined in (19). The sufficiency of (C1) for (127) follows from: $\mathbf{0} \leq \overline{\mathbf{S}} \leq \mathbf{S} \implies \rho(\overline{\mathbf{S}}) \leq \rho(\mathbf{S})$ [25, Cor. 2.2.22]; which completes the proof. ∎

## APPENDIX D
## PROOF OF THEOREM 12

We may focus w.l.o.g. only on the case in which the channel matrices $\{\mathbf{H}_{qq}\}_{q\in\Omega}$ are full (row/column) rank matrices. The case of rank deficient matrices can be cast in that of full column-rank channel matrices, as proved in Appendix C.[13] Moreover, for the sake of notation simplicity, we consider in the following only the case in which $\mathrm{rank}(\mathbf{H}_{qq}) = n_{R_q}$ for all $q \in \Omega$ or $\mathrm{rank}(\mathbf{H}_{qq}) = n_{T_q}$ for all $q \in \Omega$. The case in which some channels $\mathbf{H}_{qq}$ are full column-rank and some others are full row-rank can be similarly addressed. The proof consists in showing that, under (27) of Theorem 5 (if $\text{rank}(\mathbf{H}_{qq}) = n_{R_q}$, $\forall q \in \Omega$) or (50) of Theorem 7 (if $\text{rank}(\mathbf{H}_{qq}) = n_{T_q}$, $\forall q \in \Omega$), conditions of the asynchronous convergence theorem in [23, Prop. 2.1] are satisfied by the asynchronous MIMO IWFA described in Algorithm 1. The asynchronous convergence theorem applied to Algorithm 1 can be restated as follows.

---

[13]It is straightforward to see that conditions for the convergence of the asynchronous MIMO IWFA applied to game $\overline{\mathscr{G}}$ defined in (126) guarantee also convergence of the asynchronous MIMO IWFA applied to the original game $\mathscr{G}$. Observe that, since the asynchronous IWFA applied to game $\mathscr{G}$ is allowed to start from any arbitrary point $\mathbf{Q}^{(0)}$ in $\mathscr{Q}$, it may happen that $\mathbf{Q}^{(0)}$ does not belong to the class of matrices defined in (124). However, all the points produced by the algorithm after the first iteration from all the users $q \in \overline{\Omega}$ as well as the Nash equilibria of $\mathscr{G}$ are confined in the class (124) for all $q \in \overline{\Omega}$.

*Theorem 18 (Asynchronous convergence theorem):* Given the waterfilling mapping $\mathbf{WF} = (\mathbf{WF}_q)_{q \in \Omega} : \mathcal{Q} \mapsto \mathcal{Q}$, with $\mathbf{WF}_q(\cdot)$ and $\mathcal{Q}_q$ defined in (15) and (6), respectively, assume that there exists a sequence of nonempty sets $\{\mathcal{X}(k)\}_k$ with

$$\ldots \subset \mathcal{X}(k+1) \subset \mathcal{X}(k) \subset \ldots \subset \mathcal{Q}, \quad (130)$$

*satisfying the next two conditions.*

C.1 (Synchronous Convergence Condition)

$$\mathbf{WF}(\mathbf{Q}) \in \mathcal{X}(k+1), \forall k \text{ and } \mathbf{Q} \in \mathcal{X}(k). \quad (131)$$

*Furthermore, if* $\{\mathbf{Q}^{(k)}\}$ *is a sequence such that* $\mathbf{Q}^{(k)} \in \mathcal{X}(k)$*, for every $k$, then every limit point of* $\{\mathbf{Q}^{(k)}\}$ *is a fixed point of* $\mathbf{WF}(\cdot)$.

C.2 (Box Condition) *For every $k$ there exist sets $\mathcal{X}_q(k) \subset \mathcal{Q}_q$ such that*

$$\mathcal{X}(k) = \mathcal{X}_1(k) \times \ldots \times \mathcal{X}_Q(k). \quad (132)$$

*Then, every limit point of* $\{\mathbf{Q}^{(k)}\}$ *generated by the asynchronous MIMO IWFA in Algorithm 1, starting from* $\mathbf{Q}^{(0)} \in \mathcal{X}(0)$*, is a fixed point of* $\mathbf{WF}(\cdot)$. □

Assume that condition (27) in Theorem 5 (if $\text{rank}(\mathbf{H}_{qq}) = n_{R_q}$, $\forall q \in \Omega$) or condition (50) in Theorem 7 (if $\text{rank}(\mathbf{H}_{qq}) = n_{T_q}$, $\forall q \in \Omega$) is satisfied, i.e.,

$$\beta = \beta(\mathbf{w}, \mathbf{S}(\mathscr{P}^\star)) \triangleq \|\mathbf{S}(\mathscr{P}^\star)\|_{\infty,\text{mat}}^{\mathbf{w}} < 1, \quad (133)$$

for some $\mathbf{w} > \mathbf{0}$, implying that the waterfilling mapping $\mathbf{WF}$ admits a unique fixed-point, denoted by $\mathbf{Q}^\star$ (cf. Theorem 9). Given

$$e_{\max}^{(0)} \triangleq \max_{\mathbf{Q}^{(0)} \in \mathcal{Q}} \|\mathbf{Q}^{(0)} - \mathbf{Q}^\star\|_{F,\text{block}}^{\mathbf{w}} < +\infty, \quad (134)$$

with $\|\cdot\|_{F,\text{block}}^{\mathbf{w}}$ given in (16), let define the candidate set $\mathcal{X}(k)$ in Theorem 18 as

$$\mathcal{X}(k) = \left\{ \mathbf{Q} \in \mathcal{Q} : \|\mathbf{Q} - \mathbf{Q}^\star\|_{F,\text{block}}^{\mathbf{w}} \leq \beta^k e_{\max}^{(0)} \right\} \subset \mathcal{Q}, \quad (135)$$

which is equal to the Cartesian product $\mathcal{X}(k) = \prod_{q=1}^{Q} \mathcal{X}_q(k)$, where

$$\mathcal{X}_q(k) \triangleq \left\{ \mathbf{Q}_q \in \mathcal{Q}_q : \|\mathbf{Q}_q - \mathbf{Q}_q^\star\|_F \leq w_q \beta^k e_{\max}^{(0)} \right\} \subset \mathcal{Q}_q. \quad (136)$$

This implies box condition **C.2**. Observe that, since $\beta < 1$, set $\mathcal{X}(k)$ in (135) satisfies also condition (130). Synchronous convergence condition **C.1** follows from the contraction property of the waterfilling mapping as proved in (26) (cf. Theorem 5) under $\beta < 1$:

$$\begin{aligned} \left\|\mathbf{Q}^{(k+1)} - \mathbf{Q}^\star\right\|_{F,\text{block}}^{\mathbf{w}} &= \left\|\mathbf{WF}(\mathbf{Q}^{(k)}) - \mathbf{Q}^\star\right\|_{F,\text{block}}^{\mathbf{w}} \\ &\leq \beta \left\|\mathbf{Q}^{(k)} - \mathbf{Q}^\star\right\|_{F,\text{block}}^{\mathbf{w}} \leq \beta^{k+1} e_{\max}^{(0)}. \end{aligned} \quad (137)$$

This implies $\mathbf{Q}^{(k+1)} \in \mathcal{X}(k+1)$ for all $k \geq 0$, whenever $\mathbf{Q}^{(k)} \in \mathcal{X}(k)$, as required in (131). Finally, the convergence to the unique fixed-point of $\mathbf{WF}(\cdot)$ of the simultaneous MIMO IWFA given by $\mathbf{Q}^{(k)} \in \mathcal{X}(k)$ for all $k$, with initial conditions $\mathbf{Q}^{(0)} \in \mathcal{X}(0) = \mathcal{Q}$, follows from the contraction property of the waterfilling mapping [23, Prop. 1.1.b].

Therefore, under (133), the asynchronous MIMO IWFA given in Algorithm 1 converges to the unique NE of game $\mathscr{G}$ for any set of initial conditions in $\mathcal{Q}$ and updating schedule. The equivalence between (133) and (C1) has already been proved in Appendix C. ■